\documentclass{article}
\usepackage{amsmath}
\usepackage{tikz}
\usepackage{graphicx}
\usepackage{float}
\usepackage{amsthm}
\usepackage{amssymb}
\usepackage{booktabs}
\usepackage{bibentry}
\usepackage{array}
\usepackage{mathrsfs}
\usepackage{bm}
\usepackage{mathtools}
\usepackage{hyperref}
\usepackage{enumerate,algorithmicx,algorithm}
\usepackage{authblk}
\usepackage{algpseudocode}
\usepackage{geometry}

\usepackage{cleveref}

\allowdisplaybreaks

\newcommand{\abs}[1]{\left|#1\right|} 

\newcommand{\ii}{\mathrm{i}}

\newtheorem{remark}{Remark}[section]

\title{An $O(\log N)$ Monte Carlo method for periodic Coulomb systems}
\author[1]{Xuanzhao Gao \thanks{xgao@flatironinstitute.org}}
\author[1]{Shidong Jiang\thanks{sjiang@flatironinstitute.org}}
\author[1,2]{Jiuyang Liang\thanks{jliang@flatironinstitute.org}}
\author[1,2]{Qi Zhou\thanks{zhouqi1729@sjtu.edu.cn}}

\affil[1]{Center for Computational Mathematics, Flatiron Institute, Simons Foundation, New York 10010,
USA}
\affil[2]{School of  Mathematical  Sciences, MOE-LSC and CMA-Shanghai, Shanghai  Jiao  Tong  University,  Shanghai 200240,  China}

\date{}

\begin{document}

\maketitle

\begin{abstract}
Efficient Monte Carlo (MC) sampling of many-body systems with long-range electrostatics is often limited by the cost of per-move energy-difference evaluation under periodic boundary conditions. We present DMK-MC, an accelerated MC method that adapts the dual-space multilevel kernel-splitting (DMK) framework to single-particle Metropolis updates. DMK-MC computes the energy change and, upon acceptance, updates the stored incoming plane-wave fields with $O(1)$ work per tree level, yielding an overall $O(\log N)$ expected work per trial move for fixed accuracy. The method decomposes the Coulomb kernel into three components: a global, periodized smooth part; a multilevel sequence of smooth difference kernels whose interactions are restricted to same-level colleague boxes; and a singular residual kernel whose short-range interactions are evaluated directly. Benchmarks on uniform, highly nonuniform, and implicit-solvent electrolyte 
and colloidal configurations  show that DMK-MC consistently outperforms a recent FMM-based $O(\log N)$ Monte Carlo method, delivering several-fold speedups at comparable tolerances.



\end{abstract}

\section{Introduction}
\label{sec::intro}

The statistical physical properties of many-body interaction models have garnered widespread interest across scientific disciplines \cite{binder1992monte,Frenkel2001Understanding,lifshitz2013statistical,poole2025accelerating}. A central computational task in studying these systems is the accurate and efficient generation of samples from the target equilibrium distribution. Two broad classes of approaches are commonly used. Molecular dynamics (MD) evolves the system by numerically integrating equations of motion, while Monte Carlo (MC) methods generate samples by stochastic exploration of phase space. Tremendous progress has been made in fast algorithms and high-performance software for MD \cite{Case2023Amber,Pall2020Heter,phillips2020NAMD,thompson2022lammps}. In contrast to MD, MC is not constrained to follow physical time evolution and can therefore propose nonphysical trial moves (e.g., large displacements and collective updates), which can substantially accelerate equilibrium sampling. Furthermore, for problems involving particle swaps, fluctuating particle numbers or charge regulation, MC methods are often a more natural modeling and sampling framework \cite{bakhshandeh2019charge,curk2021charge,liang2017gpu,wang2025chemical}.
In Metropolis--Hastings (MH) schemes \cite{hastings1970monte,metropolis1953equation}, each step modifies only a small subset of particles (often a single particle), and the proposal is accepted or rejected based on the energy difference between configurations. Consequently, provided accuracy is maintained, the overall efficiency of MC sampling is frequently dictated by how quickly and reliably one can compute these per-move energy changes.

For charged systems, the dominant bottleneck is the long-range nature of the Coulomb kernel \cite{campa2009statistical,french2010long,walker2011electrostatics}.
A single-particle move changes its interaction with all other charges, so a direct update is $O(N)$ per MH step. In addition, three-dimensional periodic boundary conditions (3D-PBC) are widely used to reduce finite-size effects \cite{Frenkel2001Understanding}. Under 3D-PBC, electrostatic energies are defined through lattice sums that are \emph{conditionally convergent}, and the physically meaningful value depends on a consistent choice of summation convention (equivalently, the treatment of the $\bm k=\bm 0$ Fourier mode/macroscopic boundary condition). Fast MC electrostatics must therefore accelerate per-move updates while enforcing periodicity and the chosen convention robustly across many accept/reject decisions.

Several families of techniques address long-range electrostatics under 3D-PBC while supporting the locality of single-particle MH moves. Ewald and Fourier-based approaches provide a principled and well-established treatment of 3D-PBC electrostatics \cite{Darden1993JCP,essmann1995smooth,Ewald1921AnnPhys,Hockney1988Computer,liang2025accelerating} and underlie many production MD/MC implementations. For MH moves, however, Fourier-grid methods are not naturally local: with incremental updates, each proposal still induces nonlocal communication and bookkeeping associated with maintaining global reciprocal-space information \cite{brukhno2021dl_monte,purton2013dl_monte}. In practice, one must balance work on the Fourier grid against direct short-range corrections: reducing the reciprocal-space workload generally requires enlarging the real-space cutoff, increasing the number of neighbors touched per move. A standard balance argument that equalizes these two contributions then suggests an $O(\sqrt{N})$ work scaling per MC proposal/update in typical regimes~\cite{kolafa1992cutoff,Saunders2021New}.

In contrast, hierarchical tree- and multipole-based methods~\cite{Barnes1986Nature,cheng1999jcp,fong2009jcp,greengard1987thesis,greengard1987fast,fmm2,fmm3,fmm4} provide a substantially more favorable route for MH dynamics: by reusing multilevel information across successive local moves, one can evaluate energy differences and apply incremental updates in $O(\log N)$ time per MH step. A series of fast MC methods based on tree/multipole representations \cite{Saunders2021New,gan2014efficient,Hoft2017fast,muller2023fast} has established a strong baseline and demonstrated that $O(\log N)$ per-move complexity is achievable in practice.

Recently, a dual-space multilevel kernel-splitting (DMK) framework has been constructed for the efficient computation of discrete and continuous convolution~\cite{jiang2025dual}. The DMK framework combines the advantages of adaptive trees, the FMM and Fourier diagonalization. It employs a multilevel splitting of the interaction kernels compatible with tree-based neighbor lists, which ensures that particle interactions are localized at all but the coarsest level. Leveraging the property that the Fourier transform can diagonalize the interaction, the framework then decouples the interactions at each level without relying on the FFT for acceleration. The framework has been implemented in an $O(N\log N)$ version and an $O(N)$ version. The $O(N)$ implementation fully utilizes bidirectional information propagation between parent and child nodes, whereas in the $O(N\log N)$ version, the contribution from each level can be computed independently. 
The framework has demonstrated performance competitive with state-of-the-art fast-summation codes for free-space problems \cite{fmm3d,malhotra2015pvfmm}.

In this paper, we develop \emph{DMK-MC}, an accelerated Monte Carlo method that modifies DMK to support the single-particle updates required by each Metropolis--Hastings (MH) move. DMK-MC evaluates energy differences and updates multilevel plane-wave representations in $O(1)$ work per level, yielding an overall $O(\log N)$ cost per MC step. DMK-MC separates \emph{trial-move evaluation} from \emph{accepted-move updates}: each proposal queries only one box per level through a precomputed incoming (local) representation, while multi-box neighborhood work is incurred only after accepted moves. This organization is advantageous in Metropolis sampling, since rejected proposals still pay the evaluation cost but avoid any incremental updates.

Compared with $O(\log N)$-per-step FMM-based MC schemes~\cite{Saunders2021New,Hoft2017fast,muller2023fast}, DMK-MC offers several additional structural advantages. First, its per-level interaction pattern is particularly simple: the multilevel plane-wave sweeping stage couples only colleague boxes, i.e., nearest neighbors at the same tree level, with the number of such boxes bounded by 27. Hierarchical multipole-style schemes, in comparison, typically maintain multiple lists and apply well-separated interactions over substantially larger neighborhoods (e.g., the interaction list, which contains about 189 boxes in 3D). Second, DMK-MC uses plane waves for both the \emph{outgoing} (multipole-like) and \emph{incoming} (local) representations, making the outgoing-to-incoming translation exceptionally simple: the incoming expansion for a target box is obtained by directly summing the outgoing expansions of its colleague source boxes. In the classic FMM~\cite{cheng1999jcp}, multipole expansions are converted  into multiple directional plane-wave expansions to accelerate multipole-to-local translations. Third, the DMK-MC framework extends naturally to other long-range, singular, nonoscillatory kernels with very mild additional overhead, whereas the most efficient FMM variants~\cite{cheng1999jcp,fmm3} typically exploit that the kernel is the fundamental solution of an underlying partial differential equation. Fourth, a practical benefit of the dual-space formulation is that conditional convergence and the macroscopic boundary convention for 3D periodic Coulomb sums can be enforced in a manner closely aligned with classical Ewald theory, through a simple adjustment of the $\bm k=\bm 0$ Fourier mode at the top level. Finally, the first two points lead to a substantially simpler implementation of DMK-MC than FMM-based MC schemes.

Extensive benchmarks demonstrate that DMK-MC consistently outperforms a recent FMM-based Monte Carlo method~\cite{Saunders2021New}. Using the authors' implementation~\cite{fmmmc} in single-core CPU tests, DMK-MC delivers several-fold speedups under identical MC simulation setups and matched target tolerances. To support reproducible evaluation, we release our software package \texttt{DMK-MC}~\cite{PDMK4MC}, which provides a C++ implementation together with a Julia interface.

The remainder of this paper is organized as follows. \Cref{sec:pswf_splitting_review} introduces the PSWF-based dual-space multilevel kernel splitting and presents our periodized smooth kernel for 3D periodic boundary conditions. \Cref{sec::MCMC} formulates Monte Carlo sampling for electrostatic systems and expresses single-particle Metropolis updates in terms of background-potential differences. In \Cref{sec::FastMC}, we present DMK-MC for efficient MC under 3D-PBC. Numerical experiments are provided in \Cref{sec::num_result}. Finally, concluding remarks are given in \Cref{sec::conclusion}.

\section{PSWF-based dual-space kernel splitting for the 3D Coulomb potential}
\label{sec:pswf_splitting_review}

In this section, we introduce the PSWF-based dual-space multilevel kernel-splitting construction
for the three-dimensional Coulomb (Laplace) kernel \cite{jiang2025dual},
\begin{equation}
K(r)=\frac{1}{r},\qquad r=\|\bm r\|,\quad \bm r\in\mathbb{R}^3,
\end{equation}
which decomposes $K$ into a global smooth component together with a hierarchy of localized
difference kernels and a short-range residual. The localized terms are compactly supported in
physical space and are evaluated efficiently via truncated Fourier representations.

\subsection{PSWF window function}
\label{subsec:pswf_windowing}

Fix a bandwidth parameter $c>0$. Define the integral operator $F_c:L^2([-1,1])\to L^2([-1,1])$ by
\begin{equation}
F_c[\varphi](x)=\int_{-1}^{1}\varphi(t)\,e^{\ii cxt}\,dt,\qquad x\in[-1,1].
\end{equation}
Let $\psi_0^{c}$ denote the eigenfunction corresponding to the largest eigenvalue $\lambda_0$ of
$F_c$, i.e.,
\begin{equation}
\int_{-1}^{1}\psi_0^{c}(t)\,e^{\ii cxt}\,dt =\lambda_0\,\psi_0^{c}(x),\qquad x\in[-1,1].
\label{eq:pswf_eig}
\end{equation}
It is known that $\lambda_0>0$ and that $\psi_0^c$ can be chosen to be strictly positive on
$(-1,1)$ and even. Here $\psi_0^c$ denotes the principal \emph{time-limited} PSWF on $[-1,1]$, and
we extend it to $\mathbb{R}$ by zero outside $[-1,1]$. With this convention, define the 1D Fourier
transform by
\begin{equation}
\widehat{\psi}_0^{\,c}(k):=\int_{\mathbb{R}}\psi_0^c(t)\,e^{\ii kt}\,dt,
\label{eq:ft_conv}
\end{equation}
so that Eq.~\eqref{eq:pswf_eig} implies the explicit identity
\begin{equation}
\widehat{\psi}_0^{\,c}(k)=\int_{-1}^{1}\psi_0^c(t)\,e^{\ii kt}\,dt=\lambda_0\,\psi_0^c(k/c),\qquad |k|\le c,
\label{eq:pswf_ft_identity}
\end{equation}
and in particular $\widehat{\psi}_0^{\,c}(0)=2\int_0^1\psi_0^c(t)\,dt$.

The accuracy parameter $\varepsilon$ enters through the choice of bandwidth $c$. We choose $c$ so
that the PSWF attains the target tolerance at the endpoint,
\begin{equation}
\psi_0^c(1)\approx\varepsilon,
\label{eq:c_eps_calib}
\end{equation}
in which case one observes the scaling
\begin{equation}
c \sim \log(1/\varepsilon),
\label{eq:c_log_eps}
\end{equation}
up to moderate constants.

\subsection{Multilevel splitting and Fourier truncation}
\label{subsec:splitting_truncation}

Let $\{h_\ell\}_{\ell\ge 0}$ denote the box sidelengths associated with a multilevel hierarchy and
assume the standard dyadic refinement $h_{\ell+1}=h_\ell/2$. Define
\begin{equation}
\phi_\ell^c(r)=\frac{1}{c_0}\int_0^{r/h_\ell}\psi_0^c(x)\,dx,
\qquad
c_0 =\int_0^1 \psi_0^c(x)\,dx =\frac12\,\widehat{\psi}_0^{\,c}(0).
\label{eq:phi_c0_def}
\end{equation}
Using $\phi_\ell^c$, define the windowed kernel $W_0$, the difference kernels
$D_\ell$, and the finest-scale residual $R_{\ell_{\max}}$ by
\begin{equation}
W_0(r)=\frac{\phi_0^c(r)}{r},\qquad
D_\ell(r)=\frac{\phi_{\ell+1}^c(r)-\phi_\ell^c(r)}{r},\qquad
R_{\ell_{\max}}(r)=\frac{1-\phi_{\ell_{\max}}^c(r)}{r}.
\label{eq:split_defs}
\end{equation}
These components satisfy the telescoping identity
\begin{equation}
  \frac{1}{r}=W_0(r)+\sum_{\ell=0}^{\ell_{\max}-1}D_\ell(r)+R_{\ell_{\max}}(r).
\label{eq:telescoping}
\end{equation}
Under the convention that $\psi_0^c$ is extended by zero outside $[-1,1]$, Eq.~\eqref{eq:phi_c0_def}
implies $\phi_\ell^c(r)=1$ for all $r\ge h_\ell$, and therefore the difference and residual
kernels are exactly compactly supported in physical space:
\begin{equation}
D_\ell(r)=0\quad \text{for all } r\ge h_\ell,
\qquad
R_{\ell_{\max}}(r)=0\quad \text{for all } r\ge h_{\ell_{\max}}.
\label{eq:exact_support}
\end{equation}

Let $\bm k\in\mathbb{R}^3$ and $k=\|\bm k\|$.
The radial Fourier transform of the difference kernel admits the explicit representation
\begin{equation}
\widehat{D}_\ell(\bm k)=\widehat{D}_\ell(k)=
\frac{4\pi}{\psi_0^c(0)}\,
\frac{\psi_0^c\!\left(k\,h_{\ell+1}/c\right)-\psi_0^c\!\left(k\,h_\ell/c\right)}{k^2},
\qquad k\neq 0,
\label{eq:Dl_hat}
\end{equation}
with the removable singularity at $k=0$ defined by 
\begin{equation}
    \widehat{D}_\ell(0) = \lim_{k\to 0}\widehat{D}_\ell(k) = \frac{2\pi \psi_0^c{''}(0) (h_{\ell+1}^2-h_\ell^2)}{c^2 \psi_0^c(0)}.
\end{equation}
$\widehat{D}_\ell$ can be truncated at a finite frequency cutoff $K_{\max}$ to achieve a
prescribed accuracy $\varepsilon$. Combining Eqs.~\eqref{eq:c_eps_calib} and \eqref{eq:Dl_hat} gives
\begin{equation}
K_{\max} = \frac{c}{h_{\ell+1}}.
\label{eq:Kmax}
\end{equation}

We choose the Fourier lattice spacing
\begin{equation}
\Delta k_{\ell} = \frac{2\pi}{3h_\ell},
\label{eq:hf_choice}
\end{equation}
following \cite{jiang2025dual}, so that the Fourier spectral approximation of $D_\ell$ is valid
uniformly on the interaction range $r\le 2h_\ell$.

Using $h_{\ell+1}=h_\ell/2$, the required truncation index is
\begin{equation}
n_f = \left\lceil \frac{K_{\max}}{\Delta k_{\ell}} \right\rceil
=\left\lceil \frac{c/h_{\ell+1}}{2\pi/(3h_\ell)} \right\rceil
=\left\lceil \frac{3c}{\pi} \right\rceil.
\label{eq:nf_closed}
\end{equation}
With $N_1:=2n_f+1$ modes per dimension, the total number of Fourier modes used for each difference
kernel is
\begin{equation}
N_f^D := N_1^3 = (2n_f+1)^3.
\label{eq:NfD}
\end{equation}
For three- and six-digit accuracy, $N_1$ is 13 and 25, respectively (cf.\ Table~3.1 in \cite{jiang2025dual}). We denote the corresponding truncated level-$\ell$ reciprocal set by
\begin{equation}
\label{eq::KDl}
\mathcal{K}_{D_\ell}
:=\{\bm{k}_\ell=\Delta k_{\ell}\bm m:\ \bm m\in\mathbb{Z}^3,\ \|\bm m\|_\infty\le n_f\},
\qquad \lvert\mathcal K_{D_\ell}\rvert=N_f^D.
\end{equation}

Finally, approximate $D_\ell$ by sampling $\widehat{D}_\ell$ on $\mathcal K_{D_\ell}$:
\begin{equation}
D_\ell(\bm r)\approx\frac{\Delta k_{\ell}^3}{(2\pi)^3}
\sum_{\bm k_\ell\in\mathcal K_{D_\ell}}
e^{\,\ii\bm k_\ell\cdot \bm r}\;\widehat{D}_\ell(\bm k_\ell).
\label{eq:fourier_trunc}
\end{equation}
The number of Fourier modes per difference kernel, $N_f^D$, is $\mathcal{O}(1)$ and
independent of $\ell$ for fixed accuracy $\varepsilon$.

\subsection{Periodized smooth kernel for periodic boundary conditions}
\label{subsec:periodized_windowed_kernel}

For periodic boundary conditions on a cubic domain $\Omega=[-L/2,L/2)^3$, the windowed kernel $W_0$
is replaced by its periodization $W_0^{\mathrm{per}}$. With $h_0=L$, define
\begin{equation}
W_0^{\mathrm{per}}(\bm r)\;=\;\sum_{\bm n\in\mathbb{Z}^3} W_0\!\bigl(\|\bm r+\bm n L\|\bigr),
\qquad \bm r\in\Omega.
\label{eq:W0_periodization}
\end{equation}
Because $W_0$ is radial, its Fourier transform is radial and admits the explicit expression
\begin{equation}
\widehat{W}_0(\bm k)=\widehat{W}_0(k)
\;=\;\frac{4\pi}{\psi_0^c(0) k^2}\,\psi_0^c\!\left(\frac{k\,h_0}{c}\right),
\qquad k\ge 0.
\label{eq:W0_hat}
\end{equation}
For the periodized kernel on $\Omega$, we use the reciprocal lattice
\begin{equation}
\bm k_{\bm m}=\frac{2\pi}{L}\bm m,\qquad \bm m\in\mathbb{Z}^3,
\end{equation}
and the Fourier series representation (with the zero mode treated consistently with the periodic
Poisson solvability condition and the chosen macroscopic boundary convention
(Eq.~\eqref{sec::zero_mode})).

Combining Eqs.~\eqref{eq:c_eps_calib} and 
\eqref{eq:W0_hat} gives the effective cutoff for the smooth periodic part,
\begin{equation}
K_{\max}^W \;=\; \frac{c}{h_0}\;=\;\frac{c}{L}.
\label{eq:KmaxW}
\end{equation}
Truncating the reciprocal lattice to $\|\bm m\|_\infty\le n_f^W$ yields
\begin{equation}
n_f^W \;=\;\left\lceil \frac{K_{\max}^W}{2\pi/L} \right\rceil
\;=\;\left\lceil \frac{c/L}{2\pi/L} \right\rceil
\;=\;\left\lceil \frac{c}{2\pi} \right\rceil,
\label{eq:nfW_closed}
\end{equation}
and we define the truncated nonzero reciprocal set
\begin{equation}
\label{eq:KW_def}
\mathcal{K}_W
:=\Bigl\{\bm k_{\bm m}=\frac{2\pi}{L}\bm m:\ \|\bm m\|_\infty\le n_f^W\Bigr\}\backslash\{\bm 0\},
\qquad \lvert\mathcal K_W\rvert = N_f^W,
\end{equation}
where
\begin{equation}
N_f^W\;=\;(2n_f^W+1)^3-1.
\end{equation}
The truncated Fourier series approximation of the periodized smooth kernel is then
\begin{equation}
W_0^{\mathrm{per}}(\bm r)\;\approx\;\frac{1}{L^3}\sum_{\bm k\in\mathcal K_W}
\widehat{W}_0(\bm k)\,e^{\,\ii\bm k\cdot \bm r}.
\label{eq:W0_fourier_trunc}
\end{equation}

It is clear that the number of Fourier modes for the smooth periodic kernel, $N_f^W$, is $\mathcal{O}(1)$
for fixed accuracy $\varepsilon$.

\section{Monte Carlo simulation in electrostatic systems}
\label{sec::MCMC}

We consider a system of $N$ charged particles with charges $\{q_i\}$ and positions $\{\bm r_i\}\subset \Omega=[-L/2,L/2)^3$, subject to 3D-PBC. We assume global charge neutrality,
\begin{equation}
\label{eq:neutrality}
\sum_{i=1}^N q_i = 0.
\end{equation}
Let $\beta=(k_BT)^{-1}$, where $k_B$ is the Boltzmann constant and $T$ is the temperature. The main purpose of MC is to sample from the Boltzmann distribution associated with an energy functional $U$,
\begin{equation}
\label{eq:target_pi}
\pi(\bm X)\propto \exp(-\beta U(\bm X)),
\end{equation}
where $\bm X=(\bm r,\bm q)=(\{\bm r_i\}_{i=1}^N,\{q_i\}_{i=1}^N)$ is the system state.

\subsection{Periodic electrostatic energy and local Metropolis updates}

The (formal) 3D-PBC Coulomb energy is given by the lattice sum
\begin{equation}
\label{eq:U_lattice}
U(\bm X)=\frac12 \sum_{i=1}^N\sum_{j=1}^N \sum_{\bm n \in \mathbb{Z}^3}{ }^{\prime}\frac{q_iq_j}{\left\|\bm r_i-\bm r_j+\bm nL\right\|},
\end{equation}
where the prime indicates exclusion of the self term $i=j$ with $\bm n=\bm 0$. Under 3D-PBC the electrostatic potential is determined only up to a convention (equivalently, a choice of the $\bm k=\bm 0$ Fourier mode / ``physical condition at infinity''); we specify the convention used in \Cref{sec::zero_mode} and treat its contribution consistently there.

To sample from Eq.~\eqref{eq:target_pi}, we employ a MH scheme. Given the current state $\bm X$, we draw a candidate state $\bm X'$ from a proposal distribution $p(\bm X'\mid \bm X)$, and accept the move $\bm X\to \bm X'$ with probability
\begin{equation}
\label{eq:mh_general}
\mathscr{P}_{\mathrm{accept}}(\bm X\to \bm X')
=\min\left\{1,\exp(-\beta \Delta U)\,\frac{p(\bm X\mid \bm X')}{p(\bm X'\mid \bm X)}\right\},
\end{equation} 
where $\Delta U := U(\bm X')-U(\bm X)$ denotes the energy difference between the current and proposed states. A common choice of $p$ is a single-particle displacement. In each MC step, we select an index
$i\in\{1,\ldots,N\}$ uniformly and propose a new position $\bm r_i'$ from a conditional proposal density
$p(\bm r_i'\mid \bm r_i)$, keeping all other particles fixed. A typical example is the symmetric uniform proposal
\begin{equation}
\label{eq:proposal}
\bm r_i' = \bm r_i + \Delta \bm r_i,\quad
\Delta \bm r_i \sim \mathrm{Unif}\!\left(B_3(\bm 0,\delta)\right),\quad
B_3(\bm 0,\delta)=\{\bm x\in\mathbb{R}^3:\|\bm x\|_2\le \delta\},
\end{equation}
followed by periodic wrapping of $\bm r_i'$ back into $\Omega$. Here, $\delta$ is tuned to achieve a desired acceptance rate. The proposed configuration is
\begin{equation}
\label{eq:proposal_state}
\bm X'=(\{\bm r_1,\ldots,\bm r_i',\ldots,\bm r_N\},\{q_{i}\}_{i=1}^{N}).
\end{equation}
When the proposal is symmetric as in Eq.~\eqref{eq:proposal}, i.e., $p(\bm r_i'\mid \bm r_i)=p(\bm r_i\mid \bm r_i')$,
Eq.~\eqref{eq:mh_general} further simplifies to the standard Metropolis rule
\begin{equation}
\label{eq:p_accept}
\mathscr{P}_{\mathrm{accept}}(\bm X\to \bm X')=\min\{1,\exp(-\beta \Delta U)\}.
\end{equation}
Therefore, the computational bottleneck of each MC step typically lies in the evaluation of $\Delta U$ for a proposed move. 

In practical MC simulations, the total potential energy may include several components such as Coulomb, Lennard-Jones (LJ), and bonded interactions when molecules are present. For \emph{short-range} terms (e.g., the residual kernel of Coulomb in our setting, LJ, and bonded interactions), inter-particle distances are evaluated using the
periodically wrapped (minimum-image) displacement. The \emph{long-range} Coulomb contribution is handled by
the periodic Fourier expansions described in Eq.~\eqref{sec::FastMC}.

\subsection{Energy difference and the background potential viewpoint}

Starting from Eq.~\eqref{eq:U_lattice}, consider a displacement move of particle $i$ from the old position $\bm r_i^{o}$ to the new position $\bm r_i^{n}$, keeping all other particles fixed. Separate the part of the lattice sum that does not involve particle $i$:
\begin{equation}
\label{eq:U_minus_i}
U_{-i}:=\frac12 \sum_{\substack{a=1\\ a\neq i}}^N\sum_{\substack{b=1\\ b\neq i}}^N \sum_{\bm n \in \mathbb{Z}^3}{ }^{\prime}\frac{q_aq_b}{\left\|\bm r_a-\bm r_b+\bm nL\right\|}.
\end{equation}
When forming the energy difference $\Delta U:=U(\bm X^{n})-U(\bm X^{o})$, all terms in $U_{-i}$ cancel. The remaining contributions are the interactions between particle $i$ and each $b\neq i$ (including all periodic images). The two cross sums with $(a=i,b)$ and $(a,b=i)$ contribute equally: relabeling indices and changing variables $\bm n\to-\bm n$ gives
$\|\bm r_b-\bm r_i+\bm nL\|=\|\bm r_i-\bm r_b-\bm nL\|$, so the factor $1/2$ is removed and only the $i$--$b$ interactions remain. Any self-image contribution associated with $(a=b=i,\bm n\neq\bm 0)$ does not depend on $\bm r_i$ and therefore cancels in a displacement move (in practice it is handled implicitly by the chosen periodic Coulomb convention, e.g.\ Ewald summation). Consequently,
\begin{equation}
\label{eq:DeltaU_background}
\Delta U
= q_i\big[\Phi_{-i}(\bm r_i^{n})-\Phi_{-i}(\bm r_i^{o})\big],
\end{equation}
where the background potential generated by all particles except $i$ is
\begin{equation}
\label{eq:Phi_minus_i}
\Phi_{-i}(\bm r):=\sum_{b\neq i} q_b \sum_{\bm n\in\mathbb Z^3}\frac{1}{\|\bm r-\bm r_b+\bm nL\|}.
\end{equation}
Although the source $\{q_b\}_{b\neq i}$ is generally non-neutral, $\Phi_{-i}$ is only used through the difference
$\Phi_{-i}(\bm r_i^{n})-\Phi_{-i}(\bm r_i^{o})$, for which any additive constant (fixed by the $\bm k=\bm 0$ convention in Ewald summation) cancels; overall charge neutrality of the full system ensures the periodic electrostatics are well posed under the same convention. Eq.~\eqref{eq:DeltaU_background} shows that each displacement move can be evaluated as a difference of the background potential at two points; this ``particle-as-target'' viewpoint is central to efficient incremental-update strategies \cite{Hoft2017fast}.

\section{An \texorpdfstring{$O(\log N)$}{O(log N)} Monte Carlo method}
\label{sec::FastMC}

\Cref{sec::MCMC} showed that a single-particle Metropolis trial for particle $i$ depends on the
energy increment $\Delta U$ in Eq.~\eqref{eq:DeltaU_background}, i.e., the background potential
$\Phi_{-i}$ evaluated at the old and proposed locations. Thus, the per-step task is to evaluate
$\Phi_{-i}$ at two points, $\bm r_i^{o}$ and $\bm r_i^{n}$, and, upon acceptance, to update the
maintained representation so that the next trial can be processed efficiently.

In this section we present DMK-MC, an $O(\log N)$-per-move algorithm for evaluating
Eq.~\eqref{eq:DeltaU_background} under 3D-PBC. DMK-MC is built on the dual-space multilevel
kernel-splitting (DMK) framework \cite{jiang2025dual}. The method maintains a global smooth periodic
field together with multilevel localized corrections on an adaptive octree, leading to the
decomposition
\begin{equation}
\label{eq:Phi_split}
\Phi_{-i}(\bm r)=\Phi_{W,-i}(\bm r)+\sum_{\ell\ge 0}\Phi_{D_\ell,-i}(\bm r)+\Phi_{R,-i}(\bm r),
\end{equation}
and hence
\begin{equation}
\label{eq:DeltaU_split}
\Delta U=\Delta U_W+\Delta U_D+\Delta U_R,
\quad
\Delta U_\star=q_i\Big(\Phi_{\star,-i}(\bm r_i^{n})-\Phi_{\star,-i}(\bm r_i^{o})\Big),
\ \ \star\in\{W,D,R\}.
\end{equation}
After an initialization stage (\Cref{sec::DMK_init}), compact per-box Fourier data are stored
so that each component of Eq.~\eqref{eq:Phi_split} can be evaluated at a particle location with $O(1)$
work per level. When a move is accepted, only the data associated with a small number of boxes along the particle's ancestor paths in the tree---the sequences of boxes containing the particle at its old and new positions, from the respective leaf boxes up to the root---and the corresponding wrapped neighbor sets of those boxes are updated.

A practical feature of DMK-MC is its fixed per-level interaction pattern. At each level, the
difference-kernel contribution couples only same-level neighbor boxes (colleagues), while the
singular short-range residual is computed directly over the standard adaptive near-neighbor set. With periodic wrapping built into these neighbor relations, both the evaluation of
$\Delta U$ and the update of stored data after acceptance require $O(1)$ work per level, yielding an
overall $O(\log N)$ cost per trial move.

\paragraph{Why incoming fields matter for Metropolis}
In DMK-MC, the long-range contribution to a single-particle trial move is evaluated by visiting, at
each level, only the ancestor box containing the target particle (i.e., the chain of parent boxes from
the particle’s leaf to the root; one box per level), since that box stores the \emph{incoming}
aggregated plane-wave data. The trial energy change also includes a short-range residual correction
computed from interactions with the List-1 neighbors of the particle’s leaf box (as defined in~\Cref{sec::Adapt_Tree}); let $n_{\mathrm{L1}}$ denote the (bounded) number of such neighbors. If the move is accepted, we perform multi-box incoming-field updates over a colleague
list of at most 27 boxes (in 3D) along both the old and new ancestor paths. Thus, rejected moves pay only the per-level incoming-field evaluation
plus a leaf-local List-1 scan, while the multi-box incoming-field updates are incurred only upon
acceptance.

For a single-particle trial, the \emph{work} scales as
\begin{equation}
\begin{aligned}
&\text{trial (propose):}\ \ O(\mathscr{L}_{\max})\ \text{(incoming-field evaluation)}\ +\ O(n_{\mathrm{L1}})\ \text{(residual)},\\
&\text{accepted update:}\ \ O(27\mathscr{L}_{\max})\ \text{(incoming-field updates)}.
\end{aligned}
\end{equation}
where $\mathscr{L}_{\max}$ is the maximum tree depth.

We now describe the adaptive octree structure and the associated neighbor sets, then present the DMK
initialization procedure and the energy-difference and update routines used during Monte Carlo
sampling.

\subsection{The adaptive octree}
\label{sec::Adapt_Tree}

We construct a level-restricted (2:1 balanced) adaptive octree on the periodic cell
$\Omega=[-L/2,L/2)^3$. Starting from the root box $R=\Omega$, any node containing more than $n_s$
particles is uniformly subdivided into 8 children; this is repeated until every leaf contains at
most $n_s$ particles. We enforce the standard 2:1 balance condition \cite{Sundar2008Bottom} so that
adjacent leaf boxes differ in level by at most one.

For a node $B$, let $\ell(B)$ denote its level (with $\ell(R)=0$). The side length of a level-$\ell$
box is
\[
h_\ell=\frac{L}{2^\ell},
\qquad
h(B)=h_{\ell(B)}.
\]
Two nodes are called \emph{adjacent} if their physical domains share a face, edge, or vertex. Under
3D-PBC, adjacency is evaluated with periodic wrapping, i.e., $B$ and $D$ are adjacent if some
periodic image of $B$ is adjacent to $D$.

We use the following neighbor sets:
\begin{itemize}
    \item \emph{Colleagues.} For a node $B$ at level $\ell$, its colleague set $\mathcal{C}(B)$
    consists of all nodes at the same level $\ell$ whose boxes are adjacent to $B$ (sharing a face,
    edge, or vertex), \emph{including $B$ itself}, with periodic wrapping. In 3D, this implies
    $\lvert\mathcal{C}(B)\rvert\le 27$.
    \item \emph{List 1 (adaptive near neighbors).} For a leaf node $B$, $\mathcal{L}_1(B)$ denotes
    the standard adaptive FMM List~1 neighbor set \cite{cheng1999jcp}, i.e., the set of leaf boxes
    adjacent to $B$ under 2:1 balance. Consequently, $\mathcal{L}_1(B)$ includes adjacent leaf boxes
    at the same level as $B$, as well as adjacent leaf boxes one level coarser or finer. In 3D, 2:1 balance yields the bound $n_{\mathrm{L1}} =|\mathcal{L}_1(B)| \le 57$.
\end{itemize}

The multilevel kernels used by DMK-MC (the periodized smooth component together with the localized
difference and residual kernels) are those constructed in
\Cref{sec:pswf_splitting_review}. In the remainder of this section, we focus on how these
kernels are paired with the tree and neighbor relations above to enable $O(1)$ work per level for
both trial evaluation and accepted-move updates.

\subsection{Energy formulation under the DMK initialization}
\label{sec::DMK_init}

We next define the data structures used by DMK-MC. The \emph{DMK initialization} constructs the
adaptive tree, fixes the Fourier mode sets associated with the prescribed tolerance, and builds the
hierarchical plane-wave information needed for energy evaluation and incremental updates during
sampling.

\paragraph{Smooth periodic energy (root level)}
Let $\mathcal{K}_W\subset \tfrac{2\pi}{L}\mathbb{Z}^3\backslash\{\bm 0\}$ be the truncated reciprocal
set used for the \emph{periodized smooth kernel} $W_0^{\mathrm{per}}$
(\Cref{subsec:periodized_windowed_kernel}). The wavevectors for this smooth periodic part lie
on the \emph{root reciprocal lattice}; we denote them by $\bm k_0\in\mathcal K_W$ to distinguish them
from the level-dependent difference-kernel wavevectors $\bm k_\ell$ introduced below.

Define the root structure factor
\begin{equation}
\label{eq::structure_factor}
\rho_R(\bm k_0):=\sum_{j=1}^N q_j\,e^{-\ii\bm k_0\cdot \bm r_j},
\qquad \bm k_0\in\mathcal K_W.
\end{equation}
Inserting the truncated Fourier series  Eq.~\eqref{eq:W0_fourier_trunc} into the smooth periodic pair energy
\begin{equation}
\label{eq::UW_def}
U_W:=\frac12\sum_{i\neq j}q_iq_j\,W_0^{\mathrm{per}}(\bm r_i-\bm r_j)
\end{equation}
yields
\begin{equation}
\label{eq::UW_expand1}
U_W\approx \frac{1}{2L^3}\sum_{\bm k_0\in\mathcal K_W}\widehat W_0(\bm k_0)
\sum_{i\neq j} q_i q_j\,e^{\ii \bm k_0\cdot(\bm r_i-\bm r_j)}.
\end{equation}
Including the diagonal terms and subtracting them back, and using that the resulting double sum factorizes, we obtain
\begin{equation}
\label{eq::UW_Fourier}
U_W
\approx \frac{1}{2L^3}\sum_{\bm k_0\in\mathcal K_W}\widehat W_0(\bm k_0)\,|\rho_R(\bm k_0)|^2
\;-\;\frac{1}{2L^3}\Big(\sum_{i=1}^N q_i^2\Big)\sum_{\bm k_0\in\mathcal K_W}\widehat W_0(\bm k_0),
\end{equation}
where the second term removes the 
$i=j$ (self-interaction) contribution.

\paragraph{Difference-kernel energies (tree-localized)}
For each level $\ell\ge 0$, the difference-kernel contribution is accumulated over \emph{non-leaf}
nodes at that level,
\begin{equation}
\label{eq::UD_separate}
U_{D_\ell}=\sum_{\substack{\ell(B)=\ell\\B\ \text{non-leaf}}}U_{D_\ell}^{B}.
\end{equation}
At level $\ell$, the Fourier modes are \emph{level-dependent}; we denote them by $\bm k_\ell$ and
they lie on the lattice with spacing $\Delta k_{\ell}$ and truncated set $\mathcal K_{D_\ell}$
(\Cref{subsec:splitting_truncation}).

A convenient boxwise form of the $D_\ell$ energy is
\begin{equation}
\label{eq::UD_boxwise_start}
U_{D_\ell}
:=\frac12\sum_{\substack{\ell(B)=\ell\\B\ \text{non-leaf}}}\ \sum_{i\in B} q_i
\sum_{S\in\mathcal C(B)}\ \sum_{\substack{j\ne i\\j\in S}} q_j\,D_\ell(\bm r_i-\bm r_j^{(\mathrm{per})}),
\end{equation}
where $\bm r_j^{(\mathrm{per})}$ denotes the periodically wrapped image consistent with the wrapped
colleague relation $S\in\mathcal C(B)$ (the colleague set $\mathcal C(B)$ includes $B$ itself).

Insert the Fourier approximation Eq.~\eqref{eq:fourier_trunc} on $\mathcal K_{D_\ell}$ and regrouping yields
\begin{equation}
\label{eq::UD_regroup_step}
U_{D_\ell}
\approx \frac{\Delta k_{\ell}^{3}}{2(2\pi)^3}
\sum_{\bm k_\ell\in\mathcal K_{D_\ell}} \widehat D_\ell(\bm k_\ell)
\sum_{\substack{\ell(B)=\ell\\B\ \text{non-leaf}}}
\Big(\sum_{i\in B} q_i e^{\ii \bm k_\ell\cdot \bm r_i}\Big)
\Big(\sum_{S\in\mathcal C(B)}\sum_{\substack{j\ne i\\j\in S}} q_j e^{-\ii \bm k_\ell\cdot \bm r_j^{(\mathrm{per})}}\Big).
\end{equation}

Define the outgoing plane-wave structure factor of a level-$\ell$ box $S$ by
\begin{equation}
\label{eq::rho_out_def}
\rho_S^{\mathrm{out}}(\bm k_\ell):=\sum_{j\in S} q_j\,e^{-\ii\bm k_\ell\cdot \bm r_j},
\qquad \bm k_\ell\in\mathcal K_{D_\ell}.
\end{equation}
If $S$ participates in $\mathcal C(B)$ through a wrapped periodic image, its contribution acquires a
phase factor. We denote the wrapped outgoing structure factor by
\begin{equation}
\label{eq::rho_out_tilde_def}
\widetilde\rho_S^{\mathrm{out}}(\bm k_\ell)
:= e^{-\ii\bm k_\ell\cdot \bm t_{S\to B}}\ \rho_S^{\mathrm{out}}(\bm k_\ell),
\end{equation}
where $\bm t_{S\to B}\in L\mathbb Z^3$ is the periodic translation used to realize the wrapped image
of $S$ adjacent to $B$ (so $\bm t_{S\to B}=\bm 0$ for the unshifted neighbor).

With this notation, the colleague aggregate in Eq.~\eqref{eq::UD_regroup_step} simplifies to
\begin{equation}
\label{eq::colleague_aggregate_identity}
\sum_{S\in\mathcal C(B)}\sum_{j\in S} q_j e^{-\ii \bm k_\ell\cdot \bm r_j^{(\mathrm{per})}}
=\sum_{S\in\mathcal C(B)} \widetilde\rho_S^{\mathrm{out}}(\bm k_\ell).
\end{equation}
We therefore define the incoming plane-wave structure factor stored on the target box $B$ as
\begin{equation}
\label{eq::incoming_planewave}
\rho_B^{\mathrm{in}}(\bm k_\ell):=\sum_{S\in\mathcal C(B)} \widetilde\rho_S^{\mathrm{out}}(\bm k_\ell).
\end{equation}
Moreover, by the definition in Eq.~\eqref{eq::rho_out_def},
\(\sum_{i\in B} q_i e^{\ii \bm k_\ell\cdot \bm r_i}=\overline{\rho_B^{\mathrm{out}}(\bm k_\ell)}\).
Substituting Eq.~\eqref{eq::colleague_aggregate_identity} and this identity into
Eq.~\eqref{eq::UD_regroup_step} yields the compact discrete form
\begin{equation}
\label{eq::UD_Calc}
U_{D_\ell}
\approx
\frac{\Delta k_{\ell}^{3}}{2(2\pi)^3}
\smashoperator{\sum_{\bm k_\ell\in\mathcal K_{D_\ell}}} \widehat D_\ell(\bm k_\ell)
\smashoperator{\sum_{\substack{\ell(B)=\ell\\B\ \text{non-leaf}}}}
\Re\Big\{\overline{\rho_B^{\mathrm{out}}(\bm k_\ell)}\ \rho_B^{\mathrm{in}}(\bm k_\ell)\Big\}
-\frac12\,D_\ell(0)\!\!\sum_{\substack{\ell(B)=\ell\\B\ \text{non-leaf}}}\sum_{i\in B} q_i^2,
\end{equation}
where $\Re\{\cdot\}$ represents the real part, and the last term corrects for excluding the diagonal contribution $j=i$ (self-interaction) in the
original summation in Eq.~\eqref{eq::UD_regroup_step}. The initialization therefore stores the incoming
plane-wave structure factors $\rho_B^{\mathrm{in}}(\bm k_\ell)$ (and uses $\rho_B^{\mathrm{out}}$ as an
intermediate quantity during accepted-move updates).

\begin{remark}
The outgoing plane-wave structure factor defined in Eq.~\eqref{eq::rho_out_def} differs from the original DMK formulation \cite{jiang2025dual} by utilizing absolute coordinates $\bm r_j$ rather than relative coordinates $\bm r_j - \bm c_S$, where $\bm c_S$ denotes the box center. While this choice may theoretically induce a loss of significance or numerical instability at high precision, it simplifies the algorithm. Specifically, the diagonal translation operator required to convert outgoing factors to incoming ones is eliminated. As shown in Eq.~\eqref{eq::incoming_planewave}, $\rho^{\mathrm{in}}$ is obtained by directly summing $\widetilde{\rho}^{\mathrm{out}}$ from colleague boxes. Since $\widetilde{\rho}^{\mathrm{out}} = \rho^{\mathrm{out}}$ for the majority of boxes, the complex multiplications associated with the translation step are effectively avoided.
\end{remark}

\paragraph{Residual energy (leaf level)}
The residual component is evaluated exclusively on leaf nodes via direct summation over the adaptive List~1 neighbor set \cite{cheng1999jcp}:
\begin{equation}
\label{eq::UR_split}
U_{R} = \sum_{B \in \mathcal{L}} U_{R}^{B},
\qquad
U_{R}^{B} = \frac{1}{2}\sum_{i\in B}\ \sum_{\substack{j\neq i\\ B_j\in \mathcal L_1(B)}}
q_iq_j\,R_{\ell_j}\!\left(\|\bm r_i-\bm r_j^{(\mathrm{per})}\|\right),
\end{equation}
where $\mathcal{L}$ denotes the set of leaf boxes, and $\bm r_j^{(\mathrm{per})}$ represents the periodically wrapped image of particle $j$ associated with the neighboring leaf box $B_j\in\mathcal L_1(B)$.

\paragraph{Initialization (data and complexity)}
The initialization phase stores: (i)~the root structure factor $\rho_R(\bm k_0)$ for $\bm k_0\in\mathcal K_W$; (ii)~the incoming plane-wave structure factors $\rho_B^{\mathrm{in}}(\bm k_\ell)$ for all non-leaf nodes $B$ on their level-specific mode sets $\mathcal K_{D_\ell}$; and (iii)~the particle-to-leaf affiliations. During sampling, DMK-MC uses only these stored fields and lists.

The initialization procedure is summarized in \Cref{alg::DMK_init}. While a standard adaptive octree approach implies a cost of $O(N \log N)$, we leverage the upward pass from the original DMK framework to compute proxy charges. This ensures each particle is processed only once, reducing the total initialization complexity to $O(N)$.
\begin{algorithm}[ht]
\caption{DMK initialization (periodized)}\label{alg::DMK_init}
\begin{algorithmic}[1]
\Require Particle configuration $\bm{X}=(\{\bm{r}_i\},\{q_i\})$; box size $L$; leaf capacity $n_s$;
tolerance $\varepsilon$.
\State Construct a 2:1 balanced adaptive octree on $\Omega$ with max $n_s$ particles per leaf.
\State Choose PSWF parameters from $\varepsilon$ and fix the Fourier mode sets $\mathcal K_W$ and
$\{\mathcal K_{D_\ell}\}_{\ell\ge 0}$ (\Cref{sec:pswf_splitting_review}).
\State Precompute and store $\widehat{W}_0(\bm k_0)$ for all $\bm k_0\in \mathcal{K}_W$ and $\widehat{D}_{\ell}(\bm k_{\ell})$ for all $\bm k_{\ell}\in \mathcal{K}_{D_\ell}$ and $\ell=0,\cdots,\mathscr{L}_{\max}$.
\State Compute $\rho_R(\bm{k})$ for $\bm{k}\in\mathcal K_W$.
\State For each level $\ell\ge 0$, compute node structure factors $\rho_B(\bm{k})$ for nodes at
level $\ell$ and build $\rho_B^{\mathrm{in}}(\bm{k})$ for each non-leaf $B$ using
$\mathcal C(B)$ with periodic wrapping.
\State Record particle indices in each leaf and store particle-to-leaf affiliations.
\Ensure Stored $\rho_R$, stored $\{\rho_B^{\mathrm{in}}\}$, leaf particle lists, and tree
connectivity.
\end{algorithmic}
\end{algorithm}

\subsection{The DMK-MC method}
\label{sec::FastMC_planewave}

We now describe how DMK-MC evaluates the Metropolis energy difference for a proposed single-particle
displacement and how it updates the stored DMK data after an accepted move. We use superscripts $o$
and $n$ for quantities associated with the old (current) and new (proposed) particle locations,
respectively.

Let particle $i$ move from $\bm r_i^o$ to $\bm r_i^n$ (after periodic wrapping of $\bm r_i^n$ back
into $\Omega$). Let $B^o$ and $B^n$ be the leaf boxes containing $\bm r_i^o$ and $\bm r_i^n$,
respectively, with levels $\ell_i^o=\ell(B^o)$ and $\ell_i^n=\ell(B^n)$. For each level $\ell$, let
$B_\ell^o$ and $B_\ell^n$ denote the (unique) level-$\ell$ ancestor boxes containing $\bm r_i^o$ and
$\bm r_i^n$.

\paragraph{Smooth periodic contribution}
Using the stored root structure factor $\rho_R(\bm k_0)$ on the reciprocal set $\mathcal K_W$
(with $\lvert\mathcal K_W\rvert=N_f^W$), the contribution $\Delta U_W$ can be evaluated in $O(N_f^W)=O(1)$ work.
A useful viewpoint is to isolate the \emph{background} structure factor that excludes particle $i$:
\begin{equation}
\label{eq::rho_minus_i}
\rho_{-i}(\bm k_0):=\sum_{j\neq i} q_j e^{-\ii \bm k_0\cdot \bm r_j}
= \rho_R(\bm k_0) - q_i e^{-\ii \bm k_0\cdot \bm r_i^o}.
\end{equation}
Then the smooth background potential (under the same zero-mode convention used in
Eq.~\eqref{eq::UW_Fourier}) is
\begin{equation}
\label{eq::PhiW_minus_i}
\Phi_{W,-i}(\bm r)=\frac{1}{L^3}\sum_{\bm k_0\in\mathcal K_W}\widehat W_0(\bm k_0)\,
\Re\{\rho_{-i}(\bm k_0)^* e^{\ii \bm k_0\cdot \bm r}\},
\end{equation}
and $\Delta U_W=q_i(\Phi_{W,-i}(\bm r_i^n)-\Phi_{W,-i}(\bm r_i^o))$ reduces to the explicit stored-data
formula
\begin{equation}
\label{eq::Delta_UW}
\Delta U_W
= \frac{1}{L^3} \sum_{\bm{k}_0\in\mathcal{K}_{W}} \widehat{W}_0(\bm{k}_0)\,
\Re \left\{ \big(q_i e^{-\ii \bm{k}_0 \cdot \bm{r}_i^{n}} - q_i e^{-\ii \bm{k}_0 \cdot \bm{r}_i^{o}}\big)
\big(\rho_R(\bm{k}_0) - q_i e^{-\ii \bm{k}_0 \cdot \bm{r}_i^{o}}\big)^* \right\}.
\end{equation}
Equivalently, one may view this as the rank-one update
\begin{equation}
\label{eq::root_rank_one_update}
\rho_R(\bm k_0)\leftarrow \rho_R(\bm k_0)+q_i\big(e^{-\ii\bm k_0\cdot \bm r_i^n}-e^{-\ii\bm k_0\cdot \bm r_i^o}\big),
\qquad \bm k_0\in\mathcal K_W,
\end{equation}
applied to the quadratic energy Eq.~\eqref{eq::UW_Fourier}; both viewpoints are identical because
self-interactions are canceled out in Eq.~\eqref{eq::Delta_UW}.

\paragraph{Difference-kernel contribution}
At each difference level $\ell\ge 0$, the moved particle interacts with the stored incoming field of
its level-$\ell$ ancestor box. Concretely, the incoming structure factor $\rho_B^{\mathrm{in}}(\bm k_\ell)$
stored on a target box $B$ represents the colleague aggregate of sources around $B$ (including
periodic wrapping), and thus determines the local background field induced by $D_\ell$ within $B$.
We define the single-particle structure factor by $\rho_i(\bm k_\ell;\bm r):=q_i e^{-\ii \bm k_\ell\cdot \bm r}$.
For a point $\bm r$ contained in a (non-leaf) level-$\ell$ box $B_\ell(\bm r)$, the background
contribution from $D_\ell$ can be evaluated from stored incoming data as
\begin{equation}
\label{eq::PhiDl_minus_i}
\Phi_{D_\ell,-i}(\bm r)\approx
\frac{\Delta k_{\ell}^{3}}{(2\pi)^3}
\sum_{\bm k_\ell\in\mathcal K_{D_\ell}} \widehat D_\ell(\bm k_\ell)\,
\Re\!\left\{ e^{\ii \bm k_\ell\cdot \bm r}\,
\Big(\rho_{B_\ell(\bm r)}^{\mathrm{in}}(\bm k_\ell)-\mathbf 1_\ell(\bm r)\,\rho_i(\bm k_\ell;\bm r_i^o)^{(\mathrm{per},\,\bm r)}\Big)\right\}.
\end{equation}
Here $\mathbf 1_\ell(\bm r)\in\{0,1\}$ indicates whether the old level-$\ell$ source box $B_\ell^o$
lies in $\{B_\ell(\bm r)\}\cup \mathcal C(B_\ell(\bm r))$ under periodic wrapping. The term
$\rho_i(\bm k_\ell;\bm r_i^o)^{(\mathrm{per},\,\bm r)}$ is the contribution of particle $i$ to the stored
incoming field of $B_\ell(\bm r)$ when it is present, i.e.,
\begin{equation}
\label{eq::self_term_explicit}
\rho_i(\bm k_\ell;\bm r_i^o)^{(\mathrm{per},\,\bm r)}
:= e^{-\ii \bm k_\ell\cdot \bm t_{B_\ell^o\to B_\ell(\bm r)}}\, q_i e^{-\ii \bm k_\ell\cdot \bm r_i^o},
\end{equation}
where $\bm t_{B_\ell^o\to B_\ell(\bm r)}\in L\mathbb Z^3$ is the periodic translation associated with
the wrapped image of $B_\ell^o$ used when forming the colleague sum for $B_\ell(\bm r)$ (and
$\bm t=\bm 0$ for the unshifted neighbor). This explicit self-removal is essential because
$\mathcal C(B)$ includes $B$ itself, so the incoming field contains same-box contributions.

Writing the per-level energy contributions directly in terms of the stored incoming fields gives
\begin{align}
U_{D_\ell}^{o}
&=
\frac{\Delta k_{\ell}^{3}}{2(2\pi)^3}
\sum_{\bm{k}_\ell\in \mathcal{K}_{D_\ell}} \widehat{D}_{\ell}(\bm{k}_\ell)\,
\Re\Big\{\rho_i(\bm k_\ell;\bm r_i^o)^*
\big(\rho_{B_\ell^o}^{\mathrm{in}}(\bm k_\ell)-\rho_i(\bm k_\ell;\bm r_i^o)\big)\Big\},
\label{eq::U_diff_o}\\
U_{D_\ell}^{n}
&=
\frac{\Delta k_{\ell}^{3}}{2(2\pi)^3}
\sum_{\bm{k}_\ell\in \mathcal{K}_{D_\ell}} \widehat{D}_{\ell}(\bm{k}_\ell)\,
\Re\Big\{\rho_i(\bm k_\ell;\bm r_i^n)^*
\big(\rho_{B_\ell^n}^{\mathrm{in}}(\bm k_\ell)-\mathbf 1_\ell\,\rho_i(\bm k_\ell;\bm r_i^o)^{(\mathrm{per})}\big)\Big\},
\label{eq::U_diff_n}
\end{align}
where $\mathbf 1_\ell\in\{0,1\}$ indicates whether $B_\ell^o$ lies in
$\{B_\ell^n\}\cup\mathcal C(B_\ell^n)$ (with periodic wrapping), and
$\rho_i(\bm k_\ell;\bm r_i^o)^{(\mathrm{per})}$ denotes the periodically phase-shifted plane wave
consistent with the wrapped-image convention used in $\widetilde\rho$.

The total difference-kernel increment is then assembled with the level-dependent leaf cutoffs:
\begin{equation}
\label{eq::Delta_UD}
\Delta U_{D}=\sum_{\ell=0}^{\ell_i^n-1}U_{D_\ell}^{n}-\sum_{\ell=0}^{\ell_i^o-1}U_{D_\ell}^{o}.
\end{equation}
Since each $\mathcal K_{D_\ell}$ contains $\lvert\mathcal K_{D_\ell}\rvert=N_f^D=O(1)$ modes, the cost is
$O(N_f^D)=O(1)$ per level.

\paragraph{Residual contribution (List 1, direct)}
The residual contribution is computed by direct summation over the adaptive List~1 neighbor leaves
\cite{cheng1999jcp}. Define
\begin{align}
U_{R}^{o}
&:= q_i \sum_{\substack{j\neq i\\ B_j\in \mathcal L_1(B^{o})}}
q_j\, R_{\ell_j}\!\left(\|\bm r_i^{o}-\bm r_j^{(\mathrm{per})}\|\right),
\label{eq::UR_o}\\
U_{R}^{n}
&:= q_i \sum_{\substack{j\neq i\\ B_j\in \mathcal L_1(B^{n})}}
q_j\, R_{\ell_j}\!\left(\|\bm r_i^{n}-\bm r_j^{(\mathrm{per})}\|\right),
\label{eq::UR_n}
\end{align}
and set
\begin{equation}
\label{eq::Delta_UR}
\Delta U_R = U_R^{n}-U_R^{o}.
\end{equation}
Because each leaf contains at most $n_s=O(1)$ particles and $\lvert\mathcal L_1(B)\rvert=O(1)$ under 2:1
balance, this cost is $O(1)$.

\paragraph{Accepted-move updates (stored fields)}
Rejected moves require only evaluation of $\Delta U_W$, $\Delta U_D$, and $\Delta U_R$ from the
currently stored data. If the move is accepted, the stored representation is updated as follows.

\emph{(i) Root structure factor.} For each $\bm k_0\in\mathcal K_W$,
\begin{equation}
\label{eq::root_update_explicit}
\rho_R(\bm k_0)\ \leftarrow\ \rho_R(\bm k_0)+q_i\big(e^{-\ii\bm k_0\cdot \bm r_i^n}-e^{-\ii\bm k_0\cdot \bm r_i^o}\big).
\end{equation}

\emph{(ii) Incoming difference-level fields.} For each level $\ell$ that is traversed by the old/new
ancestor paths, only a constant-size wrapped colleague stencil of incoming fields changes. In
particular, the moved particle changes the (wrapped) source contribution associated with the level-$\ell$
ancestor box that contains it; consequently, only targets $T$ in the wrapped colleague neighborhood
of that source box require incoming-field updates. Using the same wrapped-image convention as in
Eq.~\eqref{eq::incoming_planewave}, the update is a direct add/subtract of the moved particle's
(periodized) plane-wave contribution on the relevant targets:
\begin{equation}
\label{eq::incoming_update_rule}
\rho_T^{\mathrm{in}}(\bm k_\ell)\ \leftarrow\ \rho_T^{\mathrm{in}}(\bm k_\ell)\ \pm\ \rho_i(\bm k_\ell;\cdot)^{(\mathrm{per})},
\qquad
T \in \mathcal C(\text{ancestor source box}),
\end{equation}
with one subtraction along the old ancestor path and one addition along the new ancestor path,
performed for each $\bm k_\ell\in\mathcal K_{D_\ell}$.

\emph{(iii) Leaf data and local rebalancing.} Update particle-to-leaf affiliation, update the particle
lists in $B^o$ and $B^n$, and locally restore 2:1 balance if needed (triggering the corresponding local
field rebuild on the modified neighborhood).

\Cref{alg::DMK_energy} summarizes the evaluation of $\Delta U=\Delta U_W+\Delta U_D+\Delta
U_R$. If the move is accepted, only a constant number of incoming fields per level (those on the
ancestor paths of $B^o$ and $B^n$ and their colleague neighborhoods) require updates, along with the
particle lists in the source/target leaves; see \Cref{alg::DMK_update}.

\begin{algorithm}[ht]
\caption{Energy difference evaluation in DMK-MC}\label{alg::DMK_energy}
\begin{algorithmic}[1]
\Require Stored DMK data; trial move $(q_i,\bm r_i^o)\mapsto(q_i,\bm r_i^n)$.
\State Locate the source and target leaves $B^{o},B^{n}$ containing $\bm r_i^o,\bm r_i^n$ (with
periodic wrapping), and determine their ancestor boxes $B_\ell^o,B_\ell^n$ along the tree.
\State Compute $\Delta U_W$ via Eq.~\eqref{eq::Delta_UW}.
\State For levels $\ell\ge 0$, compute $U_{D_\ell}^{o}$ and $U_{D_\ell}^{n}$ via
Eqs.~\eqref{eq::U_diff_o}--\eqref{eq::U_diff_n} using wrapped colleague membership, and form $\Delta U_D$
via Eq.~\eqref{eq::Delta_UD}.
\State Compute $\Delta U_R$ via Eqs.~\eqref{eq::UR_o}--\eqref{eq::Delta_UR}.
\Ensure $\Delta U=\Delta U_W+\Delta U_D+\Delta U_R$.
\end{algorithmic}
\end{algorithm}

\begin{algorithm}[ht]
\caption{Update of stored DMK data after an accepted move}\label{alg::DMK_update}
\begin{algorithmic}[1]
\Require Stored DMK data; accepted move $(q_i,\bm r_i^o)\mapsto(q_i,\bm r_i^n)$.
\State Update the root structure factor:
$\rho_R(\bm{k}_0)\leftarrow\rho_R(\bm{k}_0)+q_ie^{-\ii\bm{k}_0\cdot \bm{r}_i^{n}}-q_ie^{-\ii\bm{k}_0\cdot \bm{r}_i^{o}}$
for $\bm{k}_0\in \mathcal{K}_{W}$.
\State For each level $\ell\ge 0$, update the incoming fields $\rho_B^{\mathrm{in}}$ for the boxes on
the ancestor paths of $B^{o}$ and $B^{n}$ and their wrapped colleague neighborhoods, applying the
corresponding periodic phase factors where needed.
\State Update particle membership in the source/target leaf lists.
\State If necessary, locally rebalance the tree to restore the 2:1 property \cite{Sundar2008Bottom}.
\Ensure Updated DMK data.
\end{algorithmic}
\end{algorithm}

\begin{remark}
    In our implementation, tree rebalancing is performed only after $O(N)$ trial moves (i.e., approximately one Monte Carlo sweep). Since typical trial moves involve small displacements, the cumulative geometric change over $N$ single-particle updates is comparable to that of a single time step in a molecular dynamics simulation, rendering more frequent rebalancing unnecessary.
\end{remark}

\subsection{The zero-frequency mode correction}
\label{sec::zero_mode}

In the smooth periodic contribution Eq.~\eqref{eq::UW_Fourier}, the Fourier sum is taken over
$\bm{k}\in \tfrac{2\pi}{L}\mathbb{Z}^3\backslash\{\bm{0}\}$. The omitted $\bm{k}=\bm{0}$ mode is not a
benign truncation: for Coulomb interactions under periodicity the lattice sum is only conditionally
convergent, and the $\bm{k}=\bm{0}$ contribution encodes the summation convention (equivalently, the
macroscopic boundary condition). This issue is classical in Ewald-type formulations and related
Fourier-space approaches \cite{hu2014infinite,liang2026fast,liang2023random,smith1981electrostatic}.

\paragraph{Conducting (tinfoil) boundary condition}
In our numerical experiments (\Cref{sec::num_result}), we adopt the conducting (``tinfoil'')
macroscopic boundary condition, for which the macroscopic-field contribution vanishes:
\begin{equation}
U_W^{0,\text{tf}} = 0.
\end{equation}
With this choice, DMK-MC requires no additional per-move macroscopic correction beyond the
nonzero reciprocal-space modes in Eqs.~\eqref{eq::UW_Fourier} and \eqref{eq::Delta_UW}.

\paragraph{Other macroscopic boundary conditions}
Under alternative macroscopic conventions, the $\bm{k}=\bm{0}$ contribution can be written in
terms of low-order moments (e.g., the dipole), which can also be updated in $O(1)$ work per move.
We summarize the standard expressions below to make the periodic-energy convention explicit and to
facilitate use of DMK-MC under different macroscopic boundary conditions.

Formally, the missing zero-mode term associated with the smooth periodic kernel is
\begin{equation}
\label{eq::UW0}
U_W^{0}
=\frac{1}{2L^3}\sum_{i,j=1}^{N}q_iq_j\lim_{\bm{k}\to\bm{0}}
\widehat{W}_0(\bm{k})\,e^{\ii\bm{k}\cdot\bm{r}_{ij}},
\qquad
\bm r_{ij}:=\bm r_i-\bm r_j.
\end{equation}
It is easy to see that
$\widehat{W}_0(\bm k)=4\pi/k^2+O(1)$ as $\bm k\to \bm 0$ since $\psi_0^c$ is even. Thus, the zero-mode ambiguity is the
same as in Ewald-type settings. Using neutrality and symmetry, one arrives at
\begin{equation}
\label{eq::zero-mode-true}
U_W^{0}
=-\frac{\pi}{L^3}\sum_{i,j=1}^{N}q_iq_j
\lim_{\bm{k}\to\bm{0}}\frac{(\bm{k}\cdot\bm{r}_{ij})^2}{k^2}.
\end{equation}
We define the direction vector in the limit $\bm k\to\bm 0$ by $\hat{\bm k}=\lim\limits_{\bm{k}\rightarrow \bm{0}}\bm k/k$ and introduce the dipole moment
\[
\bm{\mathcal M}:=\sum_{i=1}^N q_i\bm r_i.
\]
This gives the directional form
\begin{equation}
\label{eq::UW0_directional}
U_W^{0}
=\frac{2\pi}{V}\big(\hat{\bm k}\cdot\bm{\mathcal M}\big)^2,
\qquad V=L^3,
\end{equation}
interpreted according to the prescribed limiting procedure for $\hat{\bm k}$. For an isotropic (spherical) summation convention,  $\hat{\bm k}$ averages out over the unit sphere and one obtains~\cite{hu2014infinite} 
\begin{equation}
\label{eq::sphere_slab}
U_W^{0,\text{sp}}=\frac{2\pi}{3V}\|\bm{\mathcal M}\|^2.
\end{equation}
The resulting $O(1)$ per-move correction for DMK-MC is
\begin{equation}
    \label{eq::Delta_UW0}
    \Delta U_W^{0,\text{sp}}=\frac{2\pi}{3V}\left[2q_i(\bm{r}_i^{n}-\bm{r}_i^{o})\cdot \bm{\mathcal{M}}+q_i^2\|\bm{r}_i^{n}-\bm{r}_i^{o}\|^2\right].
\end{equation}
After an accepted move, we update the dipole moment by $\bm{\mathcal{M}}\rightarrow \bm{\mathcal{M}}+q_i(\bm{r}_i^{n}-\bm{r}_i^{o})$, and keep $\bm{\mathcal M}$ unchanged if the move is rejected. For the $z$-slab summation convention, which corresponds to applying the condition $k_x=k_y=0$ before taking the limit of $k_z\rightarrow 0$, one has $\hat{\bm{k}}=(0,0,k_z/|k_z|)$ and thus  
\begin{equation}
\label{eq::slab}
U_W^{0,\text{sl}}=\frac{2\pi}{V}\mathcal M_z^2,
\qquad
\mathcal M_z:=\sum_{i=1}^N q_i z_i.
\end{equation}
The corresponding correction is $\Delta U_{W}^{0,\text{sl}}=2\pi V^{-1}\left[2q_{i}(\bm{r}_{i}^{n}-\bm{r}_{i}^{o})\cdot \bm{\mathcal{M}}+q_{i}^2|\bm{r}_{i}^{n}-\bm{r}_{i}^{o}|^2\right]$.
For other conversions, the derivations are analogous to the Ewald splitting in Refs.~\cite{hu2014infinite,smith1981electrostatic} and are omitted for brevity.

\subsection{Complexity analysis}
\label{sec::error}

We estimate the computational complexity of DMK-MC for a fixed prescribed accuracy. The dominant
work in each MC step comes from the energy-difference evaluation in \Cref{alg::DMK_energy}
and, if the move is accepted, the data-structure update in \Cref{alg::DMK_update}. Since
only one particle is displaced per step, violations of the 2:1 balance condition are infrequent;
moreover, local refinements/coarsenings needed to restore balance and to enforce the leaf capacity
$n_s$ are inexpensive \cite{Sundar2008Bottom}. We therefore focus on the costs of evaluation and
update.

Let $\mathcal C_{\mathrm{eval}}$ denote the cost of evaluating $\Delta U$ for a proposed move, and
let $\mathcal C_{\mathrm{update}}$ denote the additional cost incurred only when the move is
accepted. With acceptance probability $\mathscr{P}_{\mathrm{accept}}$, the expected cost per MC step is
\begin{equation}
\label{eq::C_expected}
\mathbb E[\mathcal C]
= \mathcal C_{\mathrm{eval}} + \mathscr{P}_{\mathrm{accept}}\,\mathcal C_{\mathrm{update}}.
\end{equation}
The evaluation stage combines the smooth periodic, difference-kernel, and residual contributions
(cf.~\eqref{eq::Delta_UW}, \eqref{eq::Delta_UD}, and \eqref{eq::Delta_UR}). Since the wavevectors lie on a lattice, the number of explicit complex exponential evaluations is minimal. Consequently, the computation of the smooth part is dominated by standard complex floating-point multiply-add arithmetic rather than expensive complex exponentials. Let $T_f$ be the cost of
one complex multiply-add (including accumulation), and let $T_s$ be the cost of one residual-kernel
pair evaluation. For fixed accuracy, $N_f^W$ and $N_f^D$ are constants independent of $N$, so
\begin{equation}
\label{eq::C_eval}
\mathcal C_{\mathrm{eval}}
=O\!\left(
N_f^W\,T_f
+\mathscr{L}_{\max}\,N_f^D\,T_f
+ 57\,n_s\,T_s
\right),
\end{equation}
where $\mathscr{L}_{\max}$ is the maximum tree depth and the constant $57$ is a geometric upper bound on the
number of leaf boxes in $\mathcal L_1(B)$ under 2:1 balance (often smaller in practice due to empty
leaves and coarse neighbors). If the move is accepted, we update (i) the root structure factor,
(ii) the plane-wave fields on each difference level for the affected boxes and their same-level
colleague neighborhoods, and (iii) the particle lists in the source/target leaves. Since a
same-level colleague neighborhood contains at most $27$ boxes (including the box itself),
\begin{equation}
\label{eq::C_update}
\mathcal C_{\mathrm{update}}
=O\!\left(
N_f^W\,T_f
+ 27\,\mathscr{L}_{\max}\,N_f^D\,T_f
\right).
\end{equation}
Storing incoming (rather than outgoing) plane-wave fields concentrates the colleague-dependent work
in the accepted-move update, so rejected moves incur only the evaluation cost
Eq.~\eqref{eq::C_eval}; this is advantageous in Metropolis sampling where rejections occur with nonzero
probability. In the quasi-uniform case, $\mathscr{L}_{\max}\approx \log_8(N/n_s)$, and for adaptive 2:1
balanced trees one still has $\mathscr{L}_{\max}=O(\log(N/n_s))$, implying $O(\log N)$ expected complexity per
MC step for fixed accuracy. 

The dominant component of the algorithm's memory footprint is the storage of incoming plane-wave structure factors for each nonleaf box. Assuming double-precision complex arithmetic (16 bytes per value), this cost is given by
\begin{equation}
    \mathcal{S}_{\mathrm{pw}} \approx 16 \cdot \frac{1}{2}\,N_B\,N_f^D,
\end{equation}
where the factor $1/2$ accounts for the conjugate symmetry inherent in the real-valued potential. For a uniform particle distribution, the number of nonleaf boxes sums to $N_B \approx N/(7n_s)$. Consequently, for three-digit accuracy ($N_f^D=13^3$), we obtain
\begin{equation}
    \mathcal{S}_{\mathrm{pw}} \approx 16 \cdot \frac{1}{2}\cdot \frac{1}{7}\cdot 13^3\,\frac{N}{n_s}
    \approx 16\cdot \frac{157}{n_s}\,N \quad \text{bytes}.
\end{equation}
With $n_s = 200$, this results in approximately $12.6$ bytes per particle. Note that doubling the number of digits requires increasing $N_f^D$, which scales $\mathcal{S}_{\mathrm{pw}}$ by a factor of approximately $8$.

Tree management incurs only modest additional overhead. For each box, we store a colleague list (bounded by 27 entries) and List~1 (bounded by 57 entries in 3D), requiring at most $84N_B$ integer indices. Additionally, we store the particle-to-leaf mapping, consisting of a particle index array of length $N$ and per-leaf offsets. The total storage may be summarized as $\mathcal{S}_{\mathrm{total}} = \mathcal{S}_{\mathrm{pw}} + \mathcal{S}_{\mathrm{list}} + \mathcal{S}_{\mathrm{assoc}}$; for the low-accuracy regimes typical of Monte Carlo calculations, this total overhead remains manageable.

\begin{remark}
Inspection of Eqs.~\eqref{eq:nf_closed} and \eqref{eq:nfW_closed} reveals that $n_f^W \approx  n_f/6$. Motivated by this observation, we modify the windowed kernel $W_0$ to incorporate the difference kernels $D_0$ and $D_1$. Specifically, the Fourier transform of $W_0$, previously defined in Eq.~\eqref{eq:W0_hat}, is redefined as
\begin{equation}
  \widehat{W}_0(\bm k) = \frac{4\pi}{\psi_0^c(0) k^2}\, \psi_0^c\! \left(\frac{k\,h_2}{c}\right) = \frac{4\pi}{\psi_0^c(0) k^2}\, \psi_0^c\! \left(\frac{kL}{4c}\right).
  \label{eq:W0_hat_modified}
\end{equation}
Consequently, the parameter $n_f^W$ in Eq.~\eqref{eq:nfW_closed} is updated to
\begin{equation}
  n_f^W \;=\;\left\lceil \frac{2c}{\pi} \right\rceil.
  \label{eq:nfW_modified}
\end{equation}
With this modification, $N_f^W$ remains smaller than $N_f^D$. This approach simplifies the implementation: starting from level 2, each box appears exactly once in the colleague list, even when accounting for periodic wrapping. Furthermore, it slightly reduces the computational cost, as the operations associated with $D_0$ and $D_1$ are effectively absorbed into the periodized smooth kernel at the root level.
\end{remark}

\section{Numerical results}
\label{sec::num_result}

In this section, we demonstrate the accuracy and efficiency of DMK-MC on three classes of electrostatic systems: (i) uniform and highly nonuniform monovalent particle systems; (ii) a $1{:}1$ NaCl electrolyte; and (iii) a colloidal-ion suspension. All experiments use the \emph{tinfoil} (conducting) convention for the $\bm k=\bm 0$ mode (\Cref{sec::zero_mode}). 
Unless otherwise specified, energy-difference accuracy is reported in terms of the geometric mean of the relative errors over $M$ independent trials,
\begin{equation}
\label{eq:err_metric}
    \varepsilon_r = \left( \prod_{1\le s\le M} \abs{ \frac{\Delta U^{(s)}_{\mathrm{DMK}}-\Delta U^{(s)}_{\mathrm{ref}}}{\Delta U^{(s)}_{\mathrm{ref}}} } \right)^{1/M},
\end{equation}
where $\Delta U_{\mathrm{ref}}$ denotes a reference energy difference computed by a higher-precision procedure specified in each experiment. Our C++ implementation, \texttt{DMK-MC}~\cite{PDMK4MC}, builds on the open-source packages \texttt{DMK}~\cite{DMKcode} and \texttt{SCTL}~\cite{SCTL}, and we also provide a Julia interface. All timings are obtained on a single CPU core of an AMD EPYC 9474F processor (1.5\,GHz) with AVX-512 support. The C++ code is compiled with GCC 13.3.0 using \texttt{-march=native -O3 -ffast-math -funroll-loops} flags.
All scripts to generate the results are available at~\cite{PDMK4MC_Benchmark}.

\subsection{Uniform and highly nonuniform systems} 
 
We first assess the accuracy and computational complexity of DMK-MC using monovalent ion systems with \emph{uniform} and \emph{highly nonuniform} particle distributions. In both cases, we initialize $N=1000$ monovalent particles ($500$  cations and $500$ anions) in a cubic box of side length $L=30$. For scaled problems, we keep the number density $\rho=N/L^3$ fixed. The two cases differ only in how particle positions are generated. In the \emph{uniform} system, particles are sampled uniformly in the box $\Omega$. 
In the \emph{highly nonuniform} system, particles are sampled uniformly on the surface of the inscribed sphere of diameter $L/2$, so the distribution is supported on a two-dimensional manifold. In building the DMK structure, the maximum number of particles per leaf node is set to $n_s=200$. All reported accuracy and timing results are averaged over $M=1000$ independent trials.

\begin{figure}[!ht]
\centering
\includegraphics[width=1.00\textwidth]{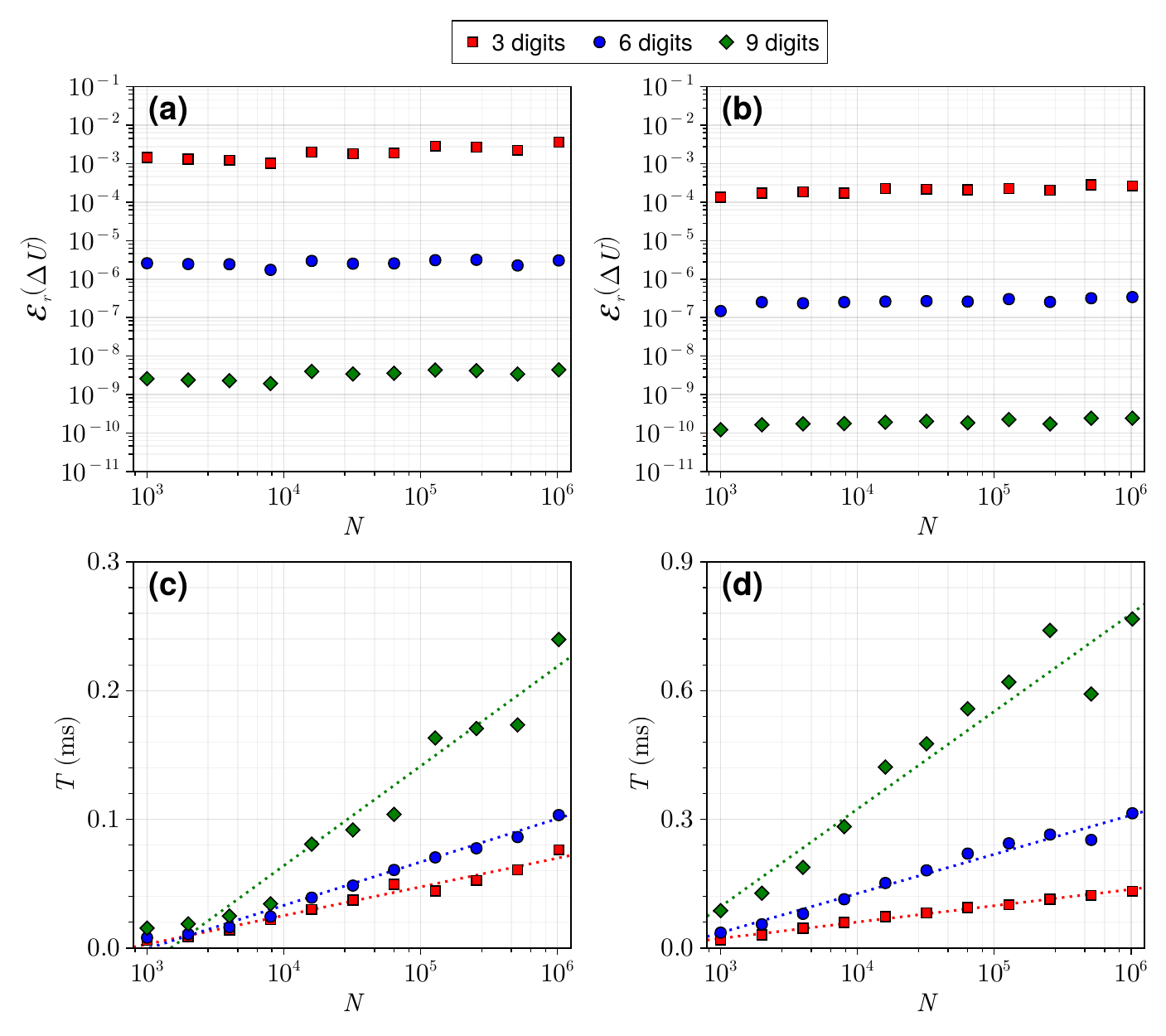}
\vspace{-0.5cm}
    \caption{Accuracy and time complexity of DMK-MC for energy-difference evaluation. (a) and (c): relative error and wall-clock time per proposal for systems with uniform distribution. (b) and (d): relative error and time per proposal for systems with highly nonuniform distribution. Dotted lines in panels (c) and (d) show least-squares fits of the form $y = a \log(x) + b$.}
    \label{fig::accuracy_scaling}
\end{figure}

\paragraph{Accuracy and $O(\log N)$ scaling}
We benchmark the energy-difference evaluation using a high-precision DMK computation with tolerance $\varepsilon=10^{-12}$ as the reference, and test prescribed tolerances $\varepsilon=10^{-3}, 10^{-6}$, and $10^{-9}$. To provide an external validation beyond DMK self-reference, we additionally compared DMK-MC energy differences against particle-mesh Ewald (PME) for the baseline case $N=1000$ at the same prescribed tolerances; the observed relative errors are consistent with DMK. 
The relative errors for uniform and highly nonuniform systems are shown in \Cref{fig::accuracy_scaling}(a--b), where we double $N$ from $1000$ up to $1{,}024 {,}000$. 
In all cases, the error remains close to the target tolerance as the system size increases. \Cref{fig::accuracy_scaling}(c--d) report the wall-clock time per energy-difference evaluation and show $O(\log N)$ scaling for both uniform and highly nonuniform distributions, consistent with the complexity analysis in \Cref{sec::error}. 
Although the two systems have the same density $\rho$, the highly nonuniform configuration typically induces a deeper 2:1 balanced tree, leading to a modestly higher per-move cost than the uniform case.

\begin{figure}[!ht]
    \centering
    \includegraphics[width=1.00\textwidth]{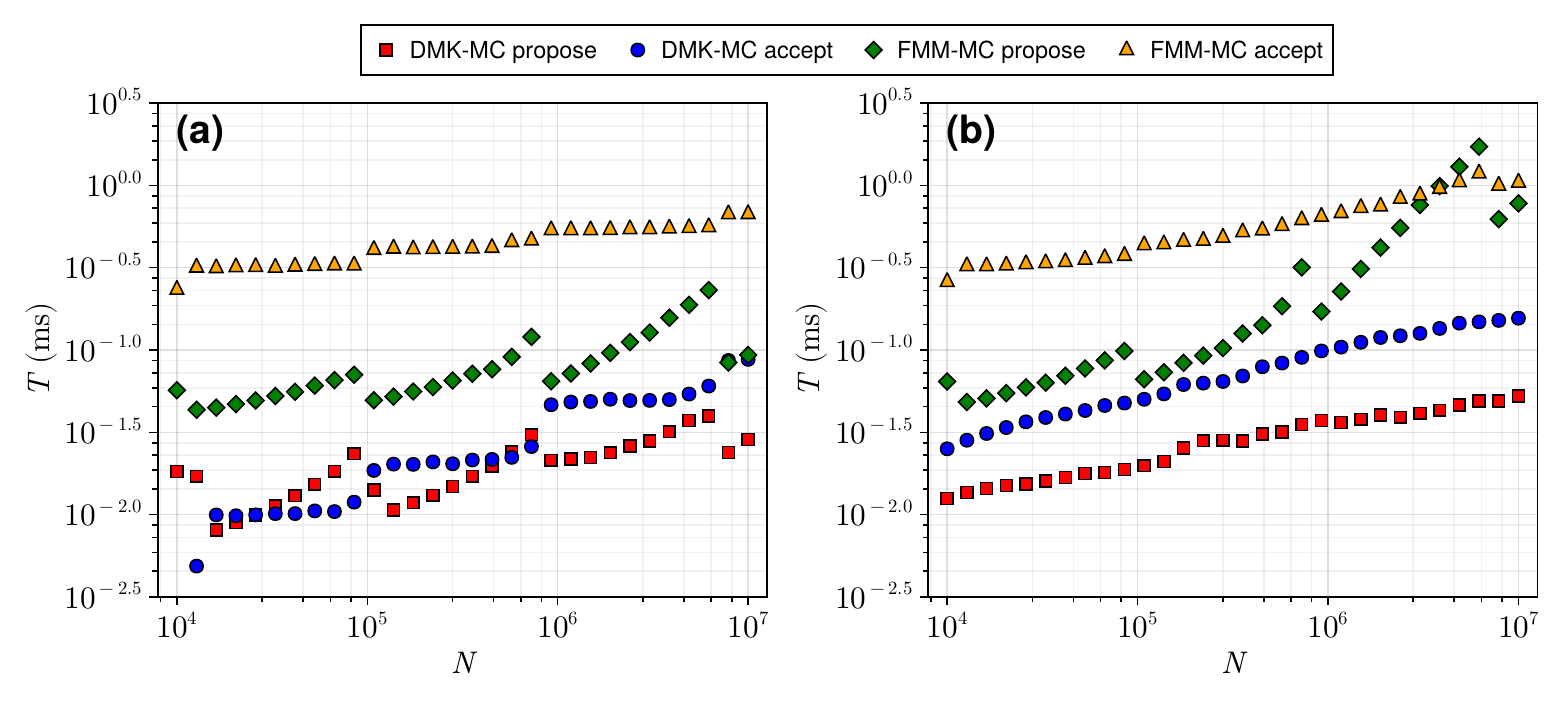}
    \vspace{-0.5cm}
    \caption{Timing comparison between DMK-MC and the FMM-based method~\cite{Saunders2021New} (log-$x$ scale), showing the per-move wall-clock time for energy-difference evaluation (``propose'') and for accepted-move updates (``accept''). The $y$-axis reports time in milliseconds. (a) Uniform distribution. (b) Highly nonuniform distribution.
}
    \label{fig::comparison_fmm_log_y}
\end{figure}

\paragraph{Comparison with an $O(\log N)$ FMM-based method}
We next compare DMK-MC with the FMM-based MC method in \cite{Saunders2021New} and its reference implementation~\cite{fmmmc} (hereafter ``FMM-MC''). All benchmarks use the same platform and simulation settings, with target accuracy $\varepsilon=10^{-3}$. The timing results for uniform and highly nonuniform distributions are shown in \Cref{fig::comparison_fmm_log_y}(a,b), where we separate the per-move cost into (i) the energy-difference evaluation for a trial move (``propose'') and (ii) the additional work to update internal data structures after an accepted move (``accept'').

For both uniform and highly nonuniform distributions, DMK-MC is faster than the FMM-MC baseline in our tests. In the uniform case, DMK-MC achieves speedups of about $4.14\times$ in the propose phase and $14.49\times$ in the accept phase; in the highly nonuniform case, the corresponding speedups are approximately $11.06\times$ (propose) and $7.32\times$ (accept). Across both distributions, the runtimes are stable, with per-trial work dominated by a bounded amount of near-field interaction and a fixed amount of same-level neighbor processing per level. The step-like changes observed in the proposal phase arise from increases in the maximum tree depth $\mathscr{L}_{\text{max}}$. As $\mathscr{L}_{\text{max}}$ increases, the number of particles per leaf node decreases, so the reduction in the last term of \Cref{eq::C_eval} can sometimes outweigh the increase in the second term of \Cref{eq::C_eval}, leading to a lower cost for evaluating $\Delta U$. This behavior is consistent with the trends reported in \cite{Saunders2021New} for FMM-MC. By contrast, the accept-phase cost is independent of the leaf capacity and therefore does not exhibit the same non-monotone behavior.

Having validated the accuracy and efficiency of the propose/accept components of DMK-MC, we next apply it to practical MC simulations of electrolyte and colloidal systems. 
In \Cref{sec::NaCl,sec::Colloidal}, we compare the sampling accuracy by comparing physical observables against that by MD simulations, and showcasing the accuracy of the per-move energy difference evaluation.

\subsection{NaCl electrolyte systems}
\label{sec::NaCl}
We study a $1{:}1$ NaCl electrolyte in the standard implicit-solvent (primitive) model. The system contains $N=1000$ mobile ions (500 $\mathrm{Na}^+$ and 500 $\mathrm{Cl}^-$) in a cubic box of side length $L=9.4~\mathrm{nm}$ with 3D-PBC. The equilibrium temperature is set as $T=300~\mathrm{K}$. Coulomb interactions are computed in a dielectric continuum with relative permittivity $\varepsilon_r=78.4$, corresponding to water at $T=300~\mathrm{K}$.
Short-range excluded-volume interactions are modeled by a LJ potential,
\begin{equation} \label{eq:lennard-jones}
    U_{\mathrm{LJ}}(r_{ij}) = 4\epsilon_{ij}\left[\left(\frac{\sigma_{ij}}{r_{ij}}\right)^{12}-\left(\frac{\sigma_{ij}}{r_{ij}}\right)^6\right],
\end{equation}
where $r_{ij}$ is the inter-particle distance, $\epsilon_{ij}$ is the well depth, and $\sigma_{ij}$ is the zero-crossing distance. We use $\sigma_{\mathrm{Na}}=0.137~\mathrm{nm}$ and $\epsilon_{\mathrm{Na}}=0.366~\mathrm{kJ/mol}$, and $\sigma_{\mathrm{Cl}}=0.251~\mathrm{nm}$ and $\epsilon_{\mathrm{Cl}}=0.149~\mathrm{kJ/mol}$ (see~\cite{PDMK4MC_Benchmark} for details). 
We use the Lorentz-Berthelot mixing rules~\cite{berthelot1898melange} for the LJ parameters, given by $\sigma_{ij}=(\sigma_i+\sigma_j)/2$ and $\epsilon_{ij}=\sqrt{\epsilon_i\epsilon_j}$, where $i,j\in\{\text{Na,Cl}\}$. 
Since $U_{\mathrm{LJ}}$ is short-ranged, its contribution to $\Delta U$ is evaluated together with the real-space residual/near-field interactions in DMK-MC. 

\begin{figure}[!ht]
    \centering   \includegraphics[width=1.00\textwidth]{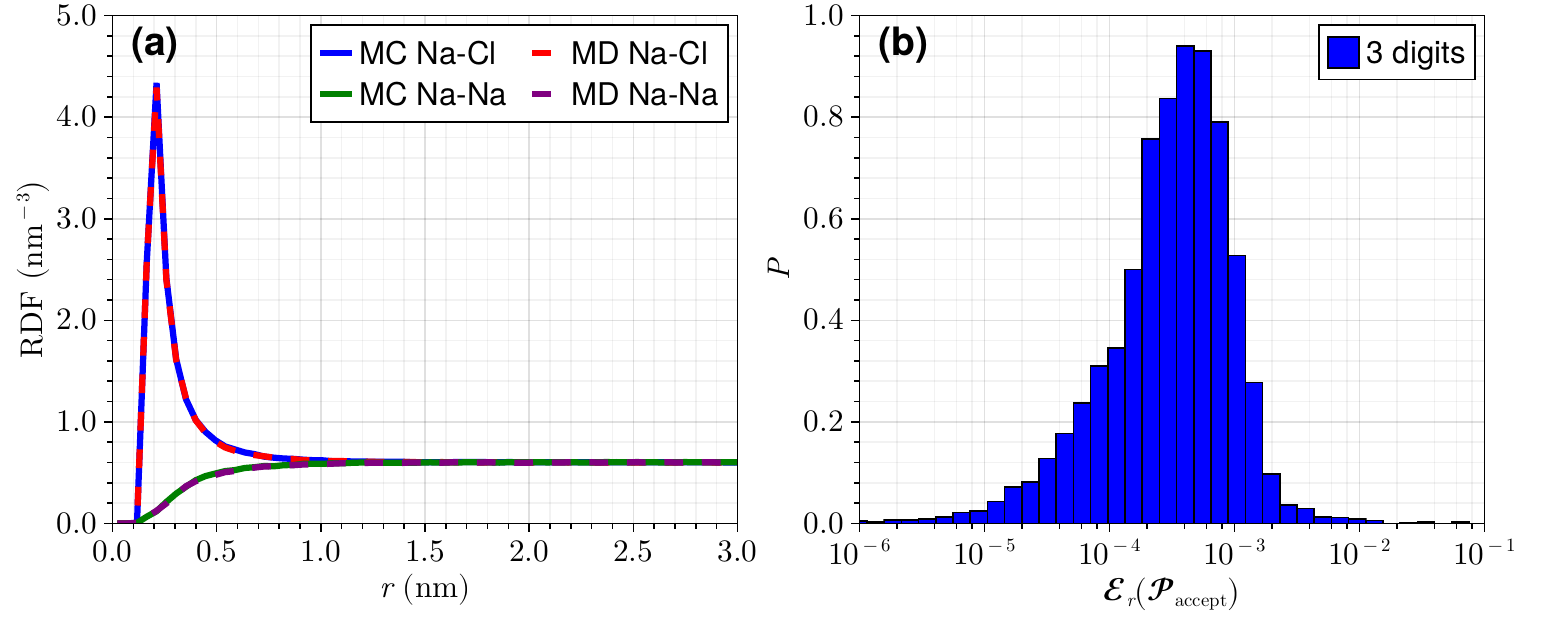}
    \vspace{-0.5cm}
    \caption{(a) Na-Na and Na-Cl RDFs for the 1:1 NaCl electrolyte. Data are shown for DMK-MC and PME-MD at the same prescribed accuracy $\varepsilon=10^{-3}$. (b) Probability profile (P) of the relative error in the acceptance probability $\mathscr{P}_{\mathrm{accept}}$ from DMK-MC at $\varepsilon=10^{-3}$, measured against a higher-accuracy reference solution. The average relative error in $\mathscr{P}_{\mathrm{accept}}$ is $6.57\times 10^{-4}$.}
    \label{fig::rdf_error_nacl}
\end{figure}

To validate sampling accuracy, we benchmark DMK-MC against an MD reference computed with PME~\cite{Darden1993JCP} (implemented in \texttt{Molly.jl}~\cite{Greener2024}). For PME-MD, we use a time step $\Delta t=1~\mathrm{fs}$, the same target accuracy $10^{-3}$, and run $5\times 10^5$ production steps after equilibration, recording configurations every $0.1~\mathrm{ps}$. This yields 5{,}000 snapshots for ensemble averaging, and the resulting MD averages are sufficiently accurate to serve as a reference. For DMK-MC, we use $n_s=50$ with tolerance $\varepsilon=10^{-3}$ and run $5\times 10^6$ production steps after equilibration (overall acceptance rate $\approx 81.6\%$). Configurations are recorded every 1{,}000 steps, again yielding 5{,}000 snapshots for ensemble averaging. In \Cref{fig::rdf_error_nacl}(a), we report the radial distribution functions (RDFs), which characterize the equilibrium structure of the electrolyte. DMK-MC reproduces the Na-Na and Na-Cl RDFs obtained from PME-MD. 
To further assess the accuracy of per-move energy-difference evaluation of DMK-MC, we show the relative errors of the acceptance probability~$\mathscr{P}_{\mathrm{accept}}$ against reference values computed by PME with accuracy of $10^{-6}$.
\Cref{fig::rdf_error_nacl}(b) shows that the resulting acceptance probability errors remain well controlled throughout the simulation, with only rare small fluctuations, consistent with the prescribed accuracy for this electrolyte system.

\subsection{Colloidal-ion systems}
\label{sec::Colloidal}

We next consider a colloidal-ion suspension as a representative highly structured system under 3D-PBC. The domain is a cubic box of side length $L=9.4~\mathrm{nm}$ containing a fixed colloid of charge $q=+100e_0$ at the origin and $900$ mobile ions (400 $\mathrm{Na}^+$ and 500 $\mathrm{Cl}^-$), so that the system is charge neutral. The solvent is treated as an aqueous dielectric continuum. In addition to Coulomb interactions, we include LJ interactions~\eqref{eq:lennard-jones} between the colloid and the mobile ions. Ion LJ parameters are the same as in \Cref{sec::NaCl}; for the colloid we set $\sigma_{\mathrm{colloid}}=1~\mathrm{nm}$ and $\epsilon_{\mathrm{colloid}}=\epsilon_{\mathrm{Na}}$.

\begin{figure}[!ht]
    \centering
    \includegraphics[width=1.00\textwidth]{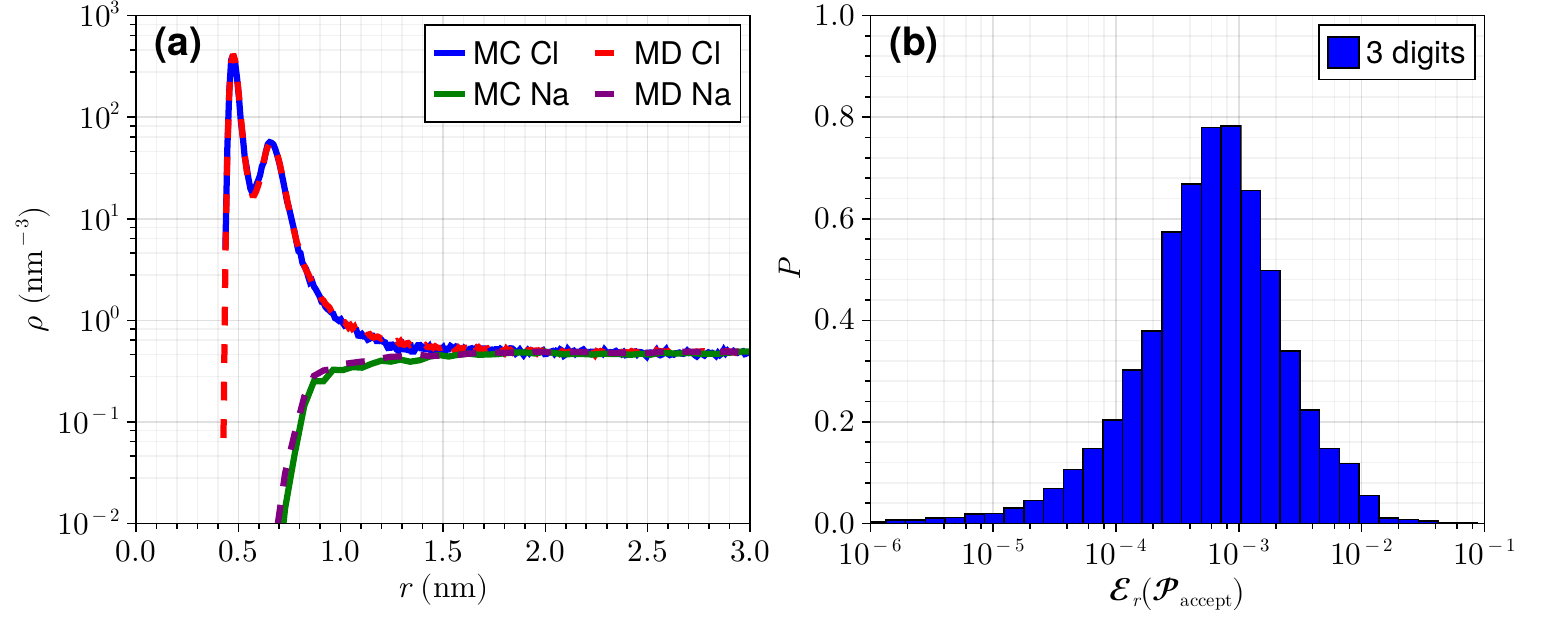}
    \vspace{-0.5cm}
    \caption{(a) Colloid--Na and colloid--Cl RDFs for a strongly inhomogeneous colloidal-ion system, and (b) probability profiles of the relative error in the acceptance probability $\mathscr{P}_{\mathrm{accept}}$ from DMK-MC at $\varepsilon=10^{-3}$, measured against a higher accuracy reference solution. The average relative error in $\mathscr{P}_{\mathrm{accept}}$ is $1.47\times 10^{-3}$.}
    \label{fig::rdf_error_colloidal}
\end{figure}

We run DMK-MC for $10^{7}$ production steps with target accuracy $\varepsilon=10^{-3}$ and leaf capacity $n_s=50$, with an overall acceptance rate $\approx 75.9\%$. 
Configurations are recorded every $1,000$ MC steps, yielding a total of $10,000$ samples. As an MD benchmark, we again use PME with target accuracy $10^{-3}$: after $10^{5}$ relaxation
steps with $\Delta t=0.1~\mathrm{fs}$, we generate a $10^{6}$-step production trajectory with $\Delta t=1~\mathrm{fs}$,
and record configurations every $100$ MD steps. This MD setup yields ensemble averages that are accurate enough to serve as a reference. \Cref{fig::rdf_error_colloidal}(a) reports the RDFs of $\mathrm{Na}^+$ and $\mathrm{Cl}^-$ relative to the colloid. The RDF profiles are more structured than in the bulk electrolyte: $\mathrm{Na}^+$ ions are depleted near the positively charged colloid, while $\mathrm{Cl}^-$ ions exhibit pronounced layering with multiple peaks. DMK-MC matches these features and agrees closely with the PME-MD references. 

We next assess the accuracy of DMK-MC energy-difference evaluations in this inhomogeneous colloidal-ion system. 
For each recorded MC trial move, we compared the acceptance probability~$\mathscr{P}_{\mathrm{accept}}$ against reference values computed by PME with accuracy of $10^{-6}$.
\Cref{fig::rdf_error_colloidal}(b) shows the distribution of relative errors of the acceptance probability over the $10{,}000$ sampled trials. Even in this challenging setting, acceptance-probability errors remain well controlled throughout the simulation. We observe a slightly heavier relative-error tail than in the uniform case, while the mean relative error in $\mathscr{P}_{\mathrm{accept}}$ remains close to $\varepsilon$. Overall, these results indicate that the chosen DMK-MC parameter settings are robust and maintain the prescribed accuracy for this highly structured colloid--ion system.

\section{Conclusion}
\label{sec::conclusion}

We have presented DMK-MC, an accelerated Monte Carlo (MC) algorithm for electrostatic systems under three-dimensional periodic boundary conditions. Building on the dual-space multilevel kernel-splitting (DMK) framework, DMK-MC decomposes the Coulomb interaction into a global periodized component and a hierarchy of localized corrections on an adaptive octree. This decomposition enables energy-difference evaluation and the associated field updates to be performed with $O(1)$ work per level, yielding an overall $O(\log N)$ cost per MC trial move for any prescribed accuracy. Comparisons with a  FMM-based Monte Carlo method~\cite{Saunders2021New,fmmmc} for 3D-PBC show that DMK-MC consistently achieves several-fold speedups.

Future work will explore extensions to reactive and grand-canonical settings (e.g., charge regulation and particle insertion/deletion), to other long-range interaction kernels, and to more general geometries such as partially periodic or anisotropic systems~\cite{gao2025fast}. Finally, DMK-MC extends naturally to other long-range, singular, nonoscillatory kernels without loss of performance, whereas FMM-based MC methods may incur additional overhead for non-PDE kernels~\cite{cheng1999jcp,fmm3}.

\section*{Acknowledgments}
The work is funded by the National Natural Science Foundation of China (grants
No. 12401570 and 125B2023). The work of J. L. was partially supported by the China Postdoctoral Science Foundation (grant No. 2024M751948).


\begin{thebibliography}{10}
\expandafter\ifx\csname url\endcsname\relax
  \def\url#1{\texttt{#1}}\fi
\expandafter\ifx\csname urlprefix\endcsname\relax\def\urlprefix{URL }\fi

\bibitem{binder1992monte}
K.~Binder, D.~W. Heermann, Monte Carlo Simulation in Statistical Physics, Vol.~8, Springer, 1992.

\bibitem{Frenkel2001Understanding}
D.~Frenkel, B.~Smit, {Understanding Molecular Simulation: From Algorithms to Applications}, Vol.~1, Elsevier, 2001.

\bibitem{lifshitz2013statistical}
E.~M. Lifshitz, L.~P. Pitaevskii, Statistical physics: theory of the condensed state, Vol.~9, Elsevier, 2013.

\bibitem{poole2025accelerating}
W.~G. Poole, M.~L. Samways, D.~Branduardi, R.~D. Taylor, M.~L. Verdonk, J.~W. Essex, Accelerating fragment-based drug discovery using grand canonical nonequilibrium candidate {Monte Carlo}, Nat. Commun. 16~(1) (2025) 6198.

\bibitem{Case2023Amber}
D.~A. Case, H.~M. Aktulga, K.~Belfon, et~al., Ambertools, J. Chem. Inf. Model. 63~(20) (2023) 6183--6191.

\bibitem{Pall2020Heter}
S.~P\'{a}ll, A.~Zhmurov, P.~Bauer, et~al., {Heterogeneous parallelization and acceleration of molecular dynamics simulations in GROMACS}, J. Chem. Phys. 153~(13) (2020) 134110.

\bibitem{phillips2020NAMD}
J.~C. Phillips, D.~J. Hardy, J.~D.~C. Maia, et~al., {Scalable molecular dynamics on CPU and GPU architectures with NAMD}, J. Chem. Phys. 153~(4) (2020) 044130.

\bibitem{thompson2022lammps}
A.~P. Thompson, H.~M. Aktulga, R.~Berger, et~al., {LAMMPS-a flexible simulation tool for particle-based materials modeling at the atomic, meso, and continuum scales}, Comput. Phys. Commun. 271 (2022) 108171.

\bibitem{bakhshandeh2019charge}
A.~Bakhshandeh, D.~Frydel, A.~Diehl, Y.~Levin, Charge regulation of colloidal particles: Theory and simulations, Phys. Rev. Lett. 123~(20) (2019) 208004.

\bibitem{curk2021charge}
T.~Curk, E.~Luijten, Charge regulation effects in nanoparticle self-assembly, Phys. Rev. Lett. 126~(13) (2021) 138003.

\bibitem{liang2017gpu}
Y.~Liang, X.~Xing, Y.~Li, A {GPU-based large-scale Monte Carlo} simulation method for systems with long-range interactions, J. Comput. Phys. 338 (2017) 252--268.

\bibitem{wang2025chemical}
Z.~Wang, T.~Yang, {Chemical order/disorder phase transitions in NiCoFeAlTiB multi-principal element alloys: A Monte Carlo analysis}, Acta Mater. 285 (2025) 120635.

\bibitem{hastings1970monte}
W.~K. Hastings, {Monte Carlo sampling methods using Markov chains and their applications}, Biometrika 57~(1) (1970) 97--109.

\bibitem{metropolis1953equation}
N.~Metropolis, A.~W. Rosenbluth, M.~N. Rosenbluth, A.~H. Teller, E.~Teller, Equation of state calculations by fast computing machines, J. Chem. Phys. 21~(6) (1953) 1087--1092.

\bibitem{campa2009statistical}
A.~Campa, T.~Dauxois, S.~Ruffo, Statistical mechanics and dynamics of solvable models with long-range interactions, Phys. Rep. 480~(3-6) (2009) 57--159.

\bibitem{french2010long}
R.~H. French, V.~A. Parsegian, R.~Podgornik, et~al., Long range interactions in nanoscale science, Rev. Mod. Phys. 82~(2) (2010) 1887--1944.

\bibitem{walker2011electrostatics}
D.~A. Walker, B.~Kowalczyk, M.~O. de~La~Cruz, B.~A. Grzybowski, Electrostatics at the nanoscale, Nanoscale 3~(4) (2011) 1316--1344.

\bibitem{Darden1993JCP}
T.~Darden, D.~York, L.~Pedersen, {Particle mesh Ewald: An N$\cdot$log(N) method for Ewald sums in large systems}, J. Chem. Phys. 98~(12) (1993) 10089--10092.

\bibitem{essmann1995smooth}
U.~Essmann, L.~Perera, M.~L. Berkowitz, T.~Darden, H.~Lee, L.~G. Pedersen, {A smooth particle mesh Ewald method}, J. Chem. Phys. 103~(19) (1995) 8577--8593.

\bibitem{Ewald1921AnnPhys}
P.~P. Ewald, {Die Berechnung optischer und elektrostatischer Gitterpotentiale}, Ann. Phys. 369~(3) (1921) 253--287.

\bibitem{Hockney1988Computer}
R.~W. Hockney, J.~W. Eastwood, {Computer Simulation Using Particles}, CRC Press, 1988.

\bibitem{liang2025accelerating}
J.~Liang, L.~Lu, A.~Barnett, L.~Greengard, S.~Jiang, Accelerating fast ewald summation with prolates for molecular dynamics simulations, arXiv preprint arXiv:2505.09727.

\bibitem{brukhno2021dl_monte}
A.~V. Brukhno, J.~Grant, T.~L. Underwood, K.~Stratford, S.~C. Parker, J.~A. Purton, N.~B. Wilding, {DL\_MONTE}: a multipurpose code for {Monte Carlo} simulation, Mol. Simul. 47~(2-3) (2021) 131--151.

\bibitem{purton2013dl_monte}
J.~Purton, J.~C. Crabtree, S.~Parker, {DL\_MONTE: a general purpose program for parallel Monte Carlo simulation}, Mol. Simul. 39~(14-15) (2013) 1240--1252.

\bibitem{kolafa1992cutoff}
J.~Kolafa, J.~W. Perram, {Cutoff errors in the Ewald summation formulae for point charge systems}, Mol. Simulat. 9~(5) (1992) 351--368.

\bibitem{Saunders2021New}
W.~R. Saunders, J.~Grant, E.~H. M\"uller, A new algorithm for electrostatic interactions in {M}onte {C}arlo simulations of charged particles, J. Comput. Phys. 430 (2021) 110099.

\bibitem{Barnes1986Nature}
J.~Barnes, P.~Hut, {A hierarchical O (N log N) force-calculation algorithm}, Nature 324~(6096) (1986) 446--449.

\bibitem{cheng1999jcp}
H.~Cheng, L.~Greengard, V.~Rokhlin, A {Fast} {Adaptive} {Multipole} {Algorithm} in {Three} {Dimensions}, J. Comput. Phys. 155~(2) (1999) 468--498.

\bibitem{fong2009jcp}
W.~Fong, E.~Darve, The black-box fast multipole method, J. Comput. Phys. 228~(23) (2009) 8712--8725.

\bibitem{greengard1987thesis}
L.~Greengard, Rapid evaluation of potential fields in particle systems, Ph.D. thesis, Yale University, New Haven, CT (USA) (1987).

\bibitem{greengard1987fast}
L.~Greengard, V.~Rokhlin, A fast algorithm for particle simulations, {J. Comput. Phys.} 73~(2) (1987) 325--348.

\bibitem{fmm2}
L.~Greengard, V.~Rokhlin, A new version of the fast multipole method for the {L}aplace equation in three dimensions, Acta Numer. 6 (1997) 229--269.

\bibitem{fmm3}
L.~Ying, G.~Biros, D.~Zorin, A kernel-independent adaptive fast multipole algorithm in two and three dimensions, J. Comput. Phys. 196~(2) (2004) 591--626.

\bibitem{fmm4}
B.~Zhang, J.~Huang, N.~P. Pitsianis, X.~Sun, A {F}ourier-series-based kernel-independent fast multipole method, J. Comput. Phys. 230~(15) (2011) 5807--5821.

\bibitem{gan2014efficient}
Z.~Gan, Z.~Xu, Efficient implementation of the {Barnes-Hut} octree algorithm for {Monte Carlo} simulations of charged systems, Sci. China Math. 57~(7) (2014) 1331--1340.

\bibitem{Hoft2017fast}
T.~A. H\"{o}ft, B.~K. Alpert, Fast updating multipole {C}oulombic potential calculation, SIAM J. Sci. Comput. 39~(3) (2017) A1038--A1061.

\bibitem{muller2023fast}
F.~M\"uller, H.~Christiansen, S.~Schnabel, W.~Janke, Fast, hierarchical, and adaptive algorithm for {Metropolis Monte Carlo} simulations of long-range interacting systems, Phys. Rev. X 13 (2023) 031006.

\bibitem{jiang2025dual}
S.~Jiang, L.~Greengard, A dual-space multilevel kernel-splitting framework for discrete and continuous convolution, Commun. Pure Appl. Math. 78~(5) (2025) 1086--1143.

\bibitem{fmm3d}
T.~Askham, Z.~Gimbutas, L.~Greengard, L.~Lu, J.~Magland, D.~Malhotra, M.~O'Neil, M.~Rachh, V.~Rokhlin, F.~Vico, {FMM3D}: Flatiron {I}nstitute {F}ast {M}ultipole {L}ibraries, \url{https://github.com/flatironinstitute/FMM3D}, release 1.0.0 (2024).

\bibitem{malhotra2015pvfmm}
D.~Malhotra, G.~Biros, {PVFMM: A parallel kernel independent FMM for particle and volume potentials}, Commun. Comput. Phys. 18~(3) (2015) 808--830.

\bibitem{fmmmc}
W.~R. Saunders, J.~Grant, E.~H. Mueller, {Code and Data Release: Monte Carlo simulations with fast and accurate electrostatics}, Zenodo (Jun. 2020).
\newline\urlprefix\url{https://doi.org/10.5281/zenodo.3873308}

\bibitem{PDMK4MC}
X.~Gao, et~al., {DMK-MC: Dual-space multilevel kernel-splitting method for {Monte Carlo} simulation}, \url{https://github.com/flatironinstitute/DMK-MC} (2025).

\bibitem{Sundar2008Bottom}
H.~Sundar, R.~S. Sampath, G.~Biros, Bottom-up construction and 2:1 balance refinement of linear octrees in parallel, SIAM J. Sci. Comput. 30~(5) (2008) 2675--2708.

\bibitem{hu2014infinite}
Z.~Hu, Infinite boundary terms of ewald sums and pairwise interactions for electrostatics in bulk and at interfaces, J. Chem. Theory Comput. 10~(12) (2014) 5254--5264.

\bibitem{liang2026fast}
J.~Liang, L.~Lu, S.~Jiang, Fast {E}wald summation with prolates for charged systems in the {NPT} ensemble, arXiv preprint arXiv:2601.00161.

\bibitem{liang2023random}
J.~Liang, Z.~Xu, Q.~Zhou, {Random batch sum-of-Gaussians method for molecular dynamics simulations of particle systems}, SIAM J. Sci. Comput. 45~(5) (2023) B591--B617.

\bibitem{smith1981electrostatic}
E.~R. Smith, Electrostatic energy in ionic crystals, Proc. R. Soc. A 375~(1763) (1981) 475--505.

\bibitem{DMKcode}
R.~Blackwell, L.~Greengard, S.~Jiang, D.~Malhotra, {DMK} {Software Library}, \url{https://github.com/flatironinstitute/dmk} (2025).

\bibitem{SCTL}
D.~Malhotra, et~al., {SCTL: Scientific Computing Template Library}, \url{https://github.com/dmalhotra/SCTL} (2022).

\bibitem{PDMK4MC_Benchmark}
X.~Gao, et~al., {DMK-MC-Benchmark}, \url{https://github.com/ArrogantGao/DMK-MC-Benchmark} (2025).

\bibitem{berthelot1898melange}
D.~Berthelot, Sur le m{\'e}lange des gaz, Compt. Rendus 126~(3) (1898) 15.

\bibitem{Greener2024}
J.~G. Greener, {Differentiable simulation to develop molecular dynamics force fields for disordered proteins}, Chem. Sci. 15 (2024) 4897--4909.

\bibitem{gao2025fast}
X.~Gao, S.~Jiang, J.~Liang, Z.~Xu, Q.~Zhou, A fast spectral sum-of-{G}aussians method for electrostatic summation in quasi-2{D} systems, Num. Math. (2025) 1--53.

\end{thebibliography}

\end{document}